\documentclass[aps,prd,floats,floatfix,nofootinbib]{revtex4-1}
\usepackage{geometry}                
\usepackage{graphicx}
\usepackage{amssymb}
\usepackage{epstopdf}
\usepackage{hyperref}

\usepackage{amsmath}
\usepackage{bbold}

\newcommand{\be}{\begin{equation}}
\newcommand{\ee}{\end{equation}}
\newcommand{\ba}{\begin{align}}
\newcommand{\ea}{\end{align}}

\usepackage{color}

\begin{document}

\title{Indirect constraints on the Georgi-Machacek model and implications for Higgs couplings}

\author{Katy Hartling}
\email{khally@physics.carleton.ca}
\author{Kunal Kumar}
\email{kkumar@physics.carleton.ca} 
\author{Heather E.~Logan}
\email{logan@physics.carleton.ca} 

\affiliation{Ottawa-Carleton Institute for Physics, Carleton University, 1125 Colonel By Drive, Ottawa, Ontario K1S 5B6, Canada}

\date{October 21, 2014}                                  

\begin{abstract}
We update the indirect constraints on the Georgi-Machacek model from $B$-physics and electroweak precision observables, including new constraints from $b \to s \gamma$ and $B^0_s \to \mu^+ \mu^-$.  We illustrate the effect of these constraints on the couplings of the Standard Model-like Higgs boson by performing scans using the most general scalar potential, subject to vacuum stability and perturbativity constraints.  We find that simultaneous enhancements of all the Higgs production cross sections by up to 39\% are still allowed after imposing these constraints.  LHC rate measurements on the Higgs pole could be blind to these enhancements if unobserved non-standard Higgs decays are present.
\end{abstract}

\maketitle 

\section{Introduction}

Since the 2012 discovery of a Standard Model (SM)-like Higgs boson at the CERN Large Hadron Collider (LHC)~\cite{Aad:2012tfa}, there has been considerable interest in models with extended Higgs sectors to be used as benchmarks for LHC searches for physics beyond the SM.  One such model is the Georgi-Machacek (GM) model~\cite{Georgi:1985nv,Chanowitz:1985ug}, which adds isospin-triplet scalar fields to the SM in a way that preserves custodial SU(2) symmetry.  Its phenomenology has been extensively studied~\cite{Gunion:1989ci,Gunion:1990dt,HHG,Haber:1999zh,Aoki:2007ah,Godfrey:2010qb,Low:2010jp,Logan:2010en,Falkowski:2012vh,Chang:2012gn, Carmi:2012in, Chiang:2012cn,Chiang:2013rua,Kanemura:2013mc,Englert:2013zpa,Killick:2013mya,
Belanger:2013xza, Englert:2013wga,Efrati:2014uta,HKL,Chiang:2014hia,Chiang:2014bia,Godunov:2014waa}.  The GM model has also been incorporated into the scalar sectors of little Higgs~\cite{Chang:2003un,Chang:2003zn} and supersymmetric~\cite{Cort:2013foa,Garcia-Pepin:2014yfa} models, and an extension with an additional isospin doublet~\cite{Hedri:2013wea} has also been considered.

The GM model has the interesting feature that the coupling strengths of the SM-like Higgs boson $h$ to $W$ or $Z$ boson pairs can be larger than in the SM.\footnote{An enhancement of the $hWW$ and $hZZ$ couplings above their SM strength while preserving custodial SU(2) symmetry can also be obtained in higher-isospin generalizations of the GM model~\cite{Galison:1983qg,Robinett:1985ec,Logan:1999if,Chang:2012gn} or in an extension of the Higgs sector by an isospin septet with appropriately-chosen hypercharge~\cite{Hisano:2013sn,Kanemura:2013mc,Alvarado:2014jva}.}  Such an enhancement is not possible in Higgs-sector extensions that contain only isospin doublets or singlets.  In light of the upcoming LHC data-taking period during which higher-precision measurements of the SM-like Higgs boson couplings and searches for additional Higgs states will be pursued, it is timely to revisit the indirect constraints on the GM model from $B$-physics and electroweak precision data.  Constraints from the oblique parameter $S$ have been studied in Refs.~\cite{Kanemura:2013mc,Englert:2013zpa,Chiang:2013rua} and constraints from the nonoblique $Z$-pole observable $R_b$ have been studied in Refs.~\cite{Haber:1999zh,Chiang:2012cn,Chiang:2013rua}.  

In this paper we point out that the dominant one-loop contributions of the additional GM Higgs bosons to nonoblique $Z$-pole observables and to $B$-physics observables can be taken over directly from calculations in the Type-I two-Higgs-doublet model (2HDM)~\cite{Branco:2011iw}.  We use this fact to determine for the first time the constraints on the GM model from $b \to s \gamma$, $B^0_s$--$\bar B^0_s$ mixing, and $B_s^0 \to \mu^+ \mu^-$.  For $b \to s \gamma$, we adapt the numerical implementation for the 2HDM in the public code SuperIso~v3.3~\cite{SuperIso}.
For $B_s^0 \to \mu^+\mu^-$, we make use of a new calculation of $B_{s}^0 \to \ell^+ \ell^-$~\cite{Li:2014fea} in the Aligned 2HDM~\cite{Pich:2009sp}.  Of these observables, we find that $b \to s \gamma$ provides the strongest constraint, though it may be surpassed in the near future as the precision on the LHC measurement of the $B_s^0 \to \mu^+ \mu^-$ branching fraction improves.
We also provide an analytic formula for the $S$ parameter in the GM model in the approximation that the new scalars are heavy compared to the $Z$ mass.  

We then examine the effect of these indirect constraints on the accessible ranges of the SM-like Higgs boson couplings.  We find that simultaneous enhancements of the $hWW$, $hZZ$, and $h \bar f f$ couplings above their SM value are still allowed, and could simultaneously enhance the SM-like Higgs boson production cross sections in all production modes by up to 39\%.  Because the LHC measures Higgs production rates only in particular Higgs-decay final states, it could be blinded to such an enhancement by the presence of new unobserved Higgs decay modes that would suppress the Higgs branching ratios into detectable final states.  Disentangling these effects will be a major phenomenological and experimental challenge at the LHC.\footnote{Of course, detecting such an enhancement in the Higgs couplings will be straightforward at a lepton-collider Higgs factory, where a direct measurement of the total Higgs production cross section in $e^+e^- \to Z h$ can be made with no reference to the Higgs decay branching ratios by using the recoil mass method; see, e.g., Ref.~\cite{Baer:2013cma}.}

This paper is organized as follows.  In Sec.~\ref{sec:model} we briefly review the GM model and set our notation.  In Sec.~\ref{sec:ST} we discuss the constraints from the oblique parameters and give our analytic formula for $S$.  In Sec.~\ref{sec:nonoblique} we discuss $R_b$ and the $B$-physics observables, and compare their constraints on the GM model parameter space.  In Sec.~\ref{sec:numerical} we illustrate the effects of these indirect constraints through numerical scans over the GM model parameter space, imposing all relevant theoretical constraints.  We conclude in Sec.~\ref{sec:conclusions}.

\section{The model}
\label{sec:model}

The scalar sector of the GM model~\cite{Georgi:1985nv,Chanowitz:1985ug} consists of the usual complex doublet $(\phi^+,\phi^0)^T$ with hypercharge\footnote{We use $Q = T^3 + Y/2$.} $Y = 1$, a real triplet $(\xi^+,\xi^0, -\xi^{+*})^T$ with $Y = 0$, and  a complex triplet $(\chi^{++},\chi^+,\chi^0)^T$ with $Y=2$.  The doublet is responsible for the fermion masses as in the SM.
Custodial symmetry is preserved at tree level by imposing a global SU(2)$_L \times$SU(2)$_R$ symmetry on the scalar potential.
In order to make this symmetry explicit, we write the doublet in the form of a bidoublet $\Phi$ and combine the triplets to form a bitriplet $X$:
\begin{equation}
	\Phi = \left( \begin{array}{cc}
	\phi^{0*} &\phi^+  \\
	-\phi^{+*} & \phi^0  \end{array} \right), \qquad
	X =
	\left(
	\begin{array}{ccc}
	\chi^{0*} & \xi^+ & \chi^{++} \\
	 -\chi^{+*} & \xi^{0} & \chi^+ \\
	 \chi^{++*} & -\xi^{+*} & \chi^0  
	\end{array}
	\right).
	\label{eq:PX}
\end{equation}
The vacuum expectation values (vevs) are defined by $\langle \Phi  \rangle = \frac{ v_{\phi}}{\sqrt{2}} \mathbb{1}_{2\times2}$  and $\langle X \rangle = v_{\chi} \mathbb{1}_{3 \times 3}$, where the $W$ and $Z$ boson masses constrain
\begin{equation}
	v_{\phi}^2 + 8 v_{\chi}^2 \equiv v^2 = \frac{1}{\sqrt{2} G_F} \approx (246~{\rm GeV})^2.
	\label{eq:vevrelation}
\end{equation} 

The most general gauge-invariant scalar potential involving these fields that conserves custodial SU(2) is given, in the conventions of Ref.~\cite{HKL}, by\footnote{A translation table to other parameterizations in the literature has been given in the appendix of Ref.~\cite{HKL}.}
\begin{eqnarray}
	V(\Phi,X) &= & \frac{\mu_2^2}{2}  \text{Tr}(\Phi^\dagger \Phi) 
	+  \frac{\mu_3^2}{2}  \text{Tr}(X^\dagger X)  
	+ \lambda_1 [\text{Tr}(\Phi^\dagger \Phi)]^2  
	+ \lambda_2 \text{Tr}(\Phi^\dagger \Phi) \text{Tr}(X^\dagger X)   \nonumber \\
          & & + \lambda_3 \text{Tr}(X^\dagger X X^\dagger X)  
          + \lambda_4 [\text{Tr}(X^\dagger X)]^2 
           - \lambda_5 \text{Tr}( \Phi^\dagger \tau^a \Phi \tau^b) \text{Tr}( X^\dagger t^a X t^b) 
           \nonumber \\
           & & - M_1 \text{Tr}(\Phi^\dagger \tau^a \Phi \tau^b)(U X U^\dagger)_{ab}  
           -  M_2 \text{Tr}(X^\dagger t^a X t^b)(U X U^\dagger)_{ab}.
           \label{eq:potential}
\end{eqnarray} 
Here the SU(2) generators for the doublet representation are $\tau^a = \sigma^a/2$ with $\sigma^a$ being the Pauli matrices, the generators for the triplet representation are
\begin{equation}
	t^1= \frac{1}{\sqrt{2}} \left( \begin{array}{ccc}
	 0 & 1  & 0  \\
	  1 & 0  & 1  \\
	  0 & 1  & 0 \end{array} \right), \qquad  
	  t^2= \frac{1}{\sqrt{2}} \left( \begin{array}{ccc}
	 0 & -i  & 0  \\
	  i & 0  & -i  \\
	  0 & i  & 0 \end{array} \right), \qquad 
	t^3= \left( \begin{array}{ccc}
	 1 & 0  & 0  \\
	  0 & 0  & 0  \\
	  0 & 0 & -1 \end{array} \right),
\end{equation}
and the matrix $U$, which rotates $X$ into the Cartesian basis, is given by~\cite{Aoki:2007ah}
\begin{equation}
	 U = \left( \begin{array}{ccc}
	- \frac{1}{\sqrt{2}} & 0 &  \frac{1}{\sqrt{2}} \\
	 - \frac{i}{\sqrt{2}} & 0  &   - \frac{i}{\sqrt{2}} \\
	   0 & 1 & 0 \end{array} \right).
	 \label{eq:U}
\end{equation}

The physical fields can be organized by their transformation properties under the custodial SU(2) symmetry into a fiveplet, a triplet, and two singlets.  The fiveplet and triplet states are given by
\begin{eqnarray}
	&&H_5^{++} = \chi^{++}, \qquad
	H_5^+ = \frac{\left(\chi^+ - \xi^+\right)}{\sqrt{2}}, \qquad
	H_5^0 = \sqrt{\frac{2}{3}} \xi^0 - \sqrt{\frac{1}{3}} \chi^{0,r}, \nonumber \\
	&&H_3^+ = - s_H \phi^+ + c_H \frac{\left(\chi^++\xi^+\right)}{\sqrt{2}}, \qquad
	H_3^0 = - s_H \phi^{0,i} + c_H \chi^{0,i},
\end{eqnarray}
where the vevs are parameterized by
\begin{equation}
	c_H \equiv \cos\theta_H = \frac{v_{\phi}}{v}, \qquad
	s_H \equiv \sin\theta_H = \frac{2\sqrt{2}\,v_\chi}{v},
\end{equation}
and we have decomposed the neutral fields into real and imaginary parts according to
\begin{equation}
	\phi^0 \to \frac{v_{\phi}}{\sqrt{2}} + \frac{\phi^{0,r} + i \phi^{0,i}}{\sqrt{2}},
	\qquad
	\chi^0 \to v_{\chi} + \frac{\chi^{0,r} + i \chi^{0,i}}{\sqrt{2}}, 
	\qquad
	\xi^0 \to v_{\chi} + \xi^0.
\end{equation}
The masses within each custodial multiplet are degenerate at tree level and can be written (after eliminating $\mu_2^2$ and $\mu_3^2$ in favor of the vevs) as\footnote{Note that the ratio $M_1/v_{\chi}$ can be written using the minimization condition $\partial V/ \partial v_{\chi} = 0$ as
\begin{equation}
	\frac{M_1}{v_{\chi}} = \frac{4}{v_{\phi}^2} 
	\left[ \mu_3^2 + (2 \lambda_2 - \lambda_5) v_{\phi}^2 
	+ 4(\lambda_3 + 3 \lambda_4) v_{\chi}^2 - 6 M_2 v_{\chi} \right],
\end{equation}
which is finite in the limit $v_{\chi} \to 0$.}
\begin{eqnarray}
	m_5^2 &=& \frac{M_1}{4 v_{\chi}} v_\phi^2 + 12 M_2 v_{\chi} 
	+ \frac{3}{2} \lambda_5 v_{\phi}^2 + 8 \lambda_3 v_{\chi}^2, \nonumber \\
	m_3^2 &=&  \frac{M_1}{4 v_{\chi}} (v_\phi^2 + 8 v_{\chi}^2) 
	+ \frac{\lambda_5}{2} (v_{\phi}^2 + 8 v_{\chi}^2) 
	= \left(  \frac{M_1}{4 v_{\chi}} + \frac{\lambda_5}{2} \right) v^2.
\end{eqnarray}

The two custodial-singlet mass eigenstates are given by
\begin{equation}
	h = \cos \alpha \, \phi^{0,r} - \sin \alpha \, H_1^{0\prime},  \qquad
	H = \sin \alpha \, \phi^{0,r} + \cos \alpha \, H_1^{0\prime},
	\label{mh-mH}
\end{equation}
where 
\begin{equation}
	H_1^{0 \prime} = \sqrt{\frac{1}{3}} \xi^0 + \sqrt{\frac{2}{3}} \chi^{0,r}.
\end{equation}
The mixing angle and masses are given by
\begin{eqnarray}
	&&\sin 2 \alpha =  \frac{2 \mathcal{M}^2_{12}}{m_H^2 - m_h^2},    \qquad
	\cos 2 \alpha =  \frac{ \mathcal{M}^2_{22} - \mathcal{M}^2_{11}  }{m_H^2 - m_h^2}, 
	\nonumber \\
	&&m^2_{h,H} = \frac{1}{2} \left[ \mathcal{M}_{11}^2 + \mathcal{M}_{22}^2
	\mp \sqrt{\left( \mathcal{M}_{11}^2 - \mathcal{M}_{22}^2 \right)^2 
	+ 4 \left( \mathcal{M}_{12}^2 \right)^2} \right],
	\label{eq:hmass}
\end{eqnarray}
where we choose $m_h < m_H$, and 
\begin{eqnarray}
	\mathcal{M}_{11}^2 &=& 8 \lambda_1 v_{\phi}^2, \nonumber \\
	\mathcal{M}_{12}^2 &=& \frac{\sqrt{3}}{2} v_{\phi} 
	\left[ - M_1 + 4 \left(2 \lambda_2 - \lambda_5 \right) v_{\chi} \right], \nonumber \\
	\mathcal{M}_{22}^2 &=& \frac{M_1 v_{\phi}^2}{4 v_{\chi}} - 6 M_2 v_{\chi} 
	+ 8 \left( \lambda_3 + 3 \lambda_4 \right) v_{\chi}^2.
\end{eqnarray}

\section{Oblique parameters}
\label{sec:ST}

The $S$ parameter~\cite{Peskin:1991sw} is given in terms of the $Z$ boson and photon self-energies as 
\begin{eqnarray}
	S &=& \frac{4 s_W^2 c_W^2}{\alpha_{\rm em} M_Z^2} \left[\Pi_{ZZ}(M_Z^2)-\Pi_{ZZ}(0)
	-\frac{c_W^2-s_W^2}{s_W c_W}\Pi_{Z\gamma}(M_Z^2)-\Pi_{\gamma\gamma}(M_Z^2)\right]
	\nonumber \\
	&\simeq& \frac{4 s_W^2 c_W^2}{\alpha_{\rm em}} \left[\Pi^{\prime}_{ZZ}(0)
	-\frac{c_W^2-s_W^2}{s_W c_W}\Pi^{\prime}_{Z\gamma}(0)-\Pi^{\prime}_{\gamma\gamma}(0)\right], 
\end{eqnarray}
where $s_W$ and $c_W$ stand for the sine and cosine of the weak mixing angle, $\alpha_{\rm em}$ is the electromagnetic fine structure constant, $M_Z$ is the $Z$ boson mass, and $\Pi^{\prime}$ denotes the derivative of the self-energy with respect to its argument $p^2$.
Here the second expression holds when the new physics scale is large compared to $M_Z$.  This second expression can be written in an analytical form, and we use it in what follows.

The new contributions to the $S$ parameter in the GM model are given by
\begin{eqnarray}
	\Delta S \equiv S_{\rm GM}-S_{\rm SM} 
	&\simeq& \frac{s_W^2 c_W^2}{\pi e^2} 
	\left\{-\frac{e^2}{12 s_W^2 c_W^2}\left(\log{m_3^2}+5\log{m_5^2}\right)
	+ 2|g_{ZhH_3^0}|^2 f_1(m_h,m_3)\right. \nonumber\\
	&& \left.+ 2|g_{ZHH_3^0}|^2\,f_1(m_H,m_3)+2\left(|g_{ZH_5^0H_3^0}|^2+2|g_{ZH_5^+H_3^{+*}}|^2\right)f_1(m_5,m_3)\right. \nonumber\\
	&& \left.+|g_{ZZh}|^2\left[\frac{f_1(M_Z,m_h)}{2 M_Z^2}-f_3(M_Z,m_h)\right]
	-|g_{ZZh}^{\rm SM}|^2\left[\frac{f_1(M_Z,m_h^{\rm SM})}{2 M_Z^2}-f_3(M_Z,m_h^{\rm SM})\right]\right. \nonumber\\
	&& \left.+|g_{ZZH}|^2\left[\frac{f_1(M_Z,m_H)}{2 M_Z^2}-f_3(M_Z,m_H)\right]
	+|g_{ZZH_5^0}|^2\left[\frac{f_1(M_Z,m_5)}{2 M_Z^2}-f_3(M_Z,m_5)\right]\right. \nonumber\\
	&& \left.+2|g_{ZW^+H_5^{+*}}|^2\left[\frac{f_1(M_W,m_5)}{2 M_W^2}-f_3(M_W,m_5)\right]\right\},
	\label{eq:SGM}
\end{eqnarray}
where $e$ is the unit of electric charge, $m_h^{\rm SM}$ is the reference SM Higgs mass for which the oblique parameters are extracted, and the loop functions are given when the new physics scale is large compared to $M_Z$ by
\begin{eqnarray}
	f_1(m_1,m_2) &=&
	\frac{1}{36(m_1^2-m_2^2)^3} \left[ 5(m_2^6-m_1^6) + 27 (m_1^4 m_2^2-m_1^2 m_2^4) 
	+ 12 (m_1^6-3 m_1^4 m_2^2) \log m_1 \right. \nonumber \\
	&& \qquad \qquad \qquad \qquad \left. + 12(3 m_1^2 m_2^4-m_2^6)\log m_2 \right]
\end{eqnarray}
and 
\begin{equation}
	f_3(m_1,m_2) 
	= \frac{m_1^4 - m_2^4 + 2 m_1^2 m_2^2 \left(\log m_2^2 - \log m_1^2\right)}
	{2 (m_1^2 - m_2^2)^3}.
\end{equation}
When their arguments are equal, $f_1$ and $f_3$ are still finite; taking $m_2^2 = m_1^2 (1 + \delta)$, where $\delta \ll 1$, $f_1$ can be expanded as
\begin{equation}
	f_1(m_1,m_2) = \frac{1}{6} \log m_1^2 + \frac{\delta}{12} + \mathcal{O}(\delta^2),
\end{equation}
and $f_3$ can be expanded as
\begin{equation}
	f_3(m_1,m_2) = \frac{1}{6 m_1^2} - \frac{\delta}{12 m_1^2} + \mathcal{O}(\delta^2/m_1^2).
\end{equation}

The couplings that appear in Eq.~(\ref{eq:SGM}) are given by~\cite{HKL}
\begin{align}
	g_{ZhH_3^0} &= -i\sqrt{\frac{2}{3}}\frac{e}{s_Wc_W}\left(\frac{s_\alpha v_\phi}{v} + \sqrt{3}\frac{c_\alpha v_\chi}{v} \right),
	&&g_{ZHH_3^0} = i\sqrt{\frac{2}{3}}\frac{e}{s_Wc_W}\left(\frac{c_\alpha v_\phi}{v} - \sqrt{3}\frac{s_\alpha v_\chi}{v} \right), \nonumber \\
	g_{ZH_5^0 H_3^0} &= -i \sqrt{\frac{1}{3}} \frac{e}{s_Wc_W}\frac{v_\phi}{v}, 
	&&g_{ZH_5^+H_3^{+*}} = \frac{e}{2 s_Wc_W}\frac{v_\phi}{v}, \nonumber \\
	g_{ZZh} &= \frac{e^2 v}{2 s_W^2 c_W^2} \left(\frac{c_\alpha v_\phi}{v} - \frac{8 s_\alpha v_\chi}{\sqrt{3} v} \right),  
	&&g_{ZZH} = \frac{e^2 v}{2 s_W^2 c_W^2} \left(\frac{s_\alpha v_\phi}{v} + \frac{8 c_\alpha v_\chi}{\sqrt{3} v} \right),  \nonumber \\
	g_{ZZ H_5^0} &= -\sqrt{\frac{8}{3}} \frac{e^2}{s_W^2 c_W^2} v_{\chi}, 
	&&g_{Z W^+ H_5^{+*}} = -\frac{\sqrt{2} e^2 v_{\chi}}{c_W s_W^2},
\end{align}
where we abbreviate $s_{\alpha} \equiv \sin\alpha$, $c_{\alpha} \equiv \cos\alpha$.
The SM coupling $g_{ZZh}^{\rm SM}$ is given by
\begin{equation}
	g_{ZZh}^{\rm SM} = \frac{e^2 v}{2 s_W^2 c_W^2}.
\end{equation}

Setting $U = 0$, the experimental values for the oblique parameters $S$ and $T$ are extracted for a reference SM Higgs mass $m_h^{\rm SM} = 125$~GeV as $S_{\rm exp} = 0.06\pm 0.09$ and $T_{\rm exp} =0.10\pm 0.07$ with a correlation coefficient of $\rho_{ST} = +0.91$~\cite{Baak:2014ora}.
We implement the constraint using a $\chi^2$ variable involving $S$ and $T$,
\begin{equation}
	\chi^2 = \sum_{i,j} (\mathcal{O}_i - \mathcal{O}_i^{\rm exp})
	(\mathcal{O}_j - \mathcal{O}_j^{\rm exp}) [\sigma^2]^{-1}_{ij},
\end{equation}
where $\mathcal{O}_i$ is the $i$th observable and $[\sigma^{2}]^{-1}_{ij}$ is the inverse of the matrix of uncertainties,
\begin{equation}
	[\sigma^2]_{ij} = \Delta \mathcal{O}_i \, \Delta \mathcal{O}_j \, \rho_{ij},
\end{equation}
where $\rho_{ij}$ are the relative correlations (note $\rho_{ii} = 1$).  For the two-observable case of interest, we can invert the matrix $\sigma^2$ explicitly and write 
\begin{equation}
	\chi^2 =  \frac{1}{\left(1-\rho_{ST}^2\right)}\left[\frac{\left(S-S_{\rm exp}\right)^2}{\left( \Delta S_{\rm exp} \right)^2}+\frac{\left(T-T_{\rm exp}\right)^2}{\left( \Delta T_{\rm exp} \right)^2}-\frac{2\,\rho_{ST}\left(S-S_{\rm exp}\right)\left(T-T_{\rm exp}\right)}{\Delta S_{\rm exp} \Delta T_{\rm exp}}\right].
\end{equation}
Here $S_{\rm exp}$ and $T_{\rm exp}$ are the experimental central values, $\Delta S_{\rm exp}$ and $\Delta T_{\rm exp}$ are their $1\sigma$ experimental uncertainties, $\rho_{ST}$ is the relative correlation between the two oblique parameters, and $S$ and $T$ are the new-physics contributions from the GM model.

It is well known that, in the GM model, hypercharge interactions break the SU(2)$_R$ global symmetry at one-loop level, yielding a divergent value for the $T$ parameter~\cite{Gunion:1990dt,Englert:2013zpa}.  This would be corrected in a more complete theory by the counterterm of an SU(2)$_R$-breaking quartic coupling in the scalar potential~\cite{Gunion:1990dt,Garcia-Pepin:2014yfa}, the finite part of which could in turn be adjusted to compensate the one-loop contributions to the $T$ parameter.  In our analysis we thus take a conservative approach and marginalize over the value of $T$ in the $\chi^2$,\footnote{In practice, we solve the constraint equation
\begin{equation}
	\left. \frac{\partial\chi^2}{\partial T} \right|_{T_{\rm min}} = 0,
\end{equation}
which yields
\begin{equation}
	T \equiv T_{\rm min}= T_{\rm exp}+\rho_{\rm ST}(S-S_{\rm exp})
		\frac{\Delta T_{\rm exp}}{\Delta S_{\rm exp}}.
\end{equation}}
resulting in a constraint on $S$ alone.
Our constraint on $S$ agrees numerically with that shown in Fig.~1 of Ref.~\cite{Chiang:2013rua}.

%
%

\section{Nonoblique and $B$-physics observables}
\label{sec:nonoblique}

Extended Higgs sectors are typically also constrained by nonoblique corrections to $Z$-pole observables, as well as $B$-physics observables.  These constraints come from one-loop diagrams involving Higgs boson couplings to fermions and to SM gauge bosons. The analysis of these constraints in the GM model is greatly simplified by the observation that the relevant diagrams are completely analogous to those of the Type-I two-Higgs-doublet model.

In the GM model, fermion masses are generated in the same way as in the SM through Yukawa couplings involving the \emph{single} SU(2)$_L$ doublet.  The resulting Feynman rules for vertices involving a scalar and two fermions, with all particles incoming, are given by~\cite{Gunion:1989ci,HHG,HKL}
\begin{eqnarray}
	h \bar f f: &\quad& -i \frac{m_f}{v} \frac{\cos \alpha}{\cos \theta_H}, \qquad \qquad
	H \bar f f: \quad -i \frac{m_f}{v} \frac{\sin \alpha}{\cos \theta_H}, \nonumber \\
	H_3^0 \bar u u: &\quad& \frac{m_u}{v} \tan \theta_H \gamma_5, \qquad \qquad
	H_3^0 \bar d d: \quad -\frac{m_d}{v} \tan \theta_H \gamma_5, \nonumber \\
	H_3^+ \bar u d: &\quad& -i \frac{\sqrt{2}}{v} V_{ud} \tan\theta_H
		\left( m_u P_L - m_d P_R \right), \nonumber \\
	H_3^+ \bar \nu \ell: &\quad& i \frac{\sqrt{2}}{v} \tan\theta_H m_{\ell} P_R.
\end{eqnarray}
Here $f$ is any charged fermion, $V_{ud}$ is the appropriate element of the Cabibbo-Kobayashi-Maskawa (CKM) matrix, and the projection operators are defined as $P_{R,L} = (1 \pm \gamma_5)/2$.  The $H_3^0 \bar \ell \ell$ couplings are the same as the $H_3^0 \bar d d$ couplings with $m_d \to m_{\ell}$. The custodial fiveplet states do not couple to fermions as they have no SU(2)$_L$-doublet component.

In particular, the scalar couplings to fermions in the GM model have exactly the same structure as those in the Type-I two-Higgs-doublet model~\cite{Branco:2011iw} with the replacement $\cot\beta \to \tan\theta_H$.  In this situation, large enhancements of scalar couplings to light fermions (in particular to the bottom quark or to charged leptons) are not possible due to perturbativity constraints on the top quark Yukawa coupling.  The dominant new-physics contributions to nonoblique $Z$-pole and $B$-physics observables are then due solely to diagrams involving scalar couplings to the top quark; in particular, diagrams involving the $H_3^+ \bar t b$ coupling.

In other words, $H_3^+$ is the only new scalar in the GM model that contributes significantly to $Z$-pole and $B$-physics observables. Since custodial symmetry requires that the $H_3^+ H_3^- Z$ coupling be identical to the $H^+H^-Z$ coupling in the 2HDM, all of the relevant $H_3^+$ couplings have the same form as those of $H^+$ in the 2HDM. This implies that all of the nonoblique $Z$-pole and $B$-physics constraints on the GM model can be obtained by making the replacements $\cot\beta \to \tan\theta_H$ and $m_{H^+} \to m_3$  in the corresponding calculations for the Type-I 2HDM.

In what follows we use this correspondence to consider the constraints on the GM model from $R_b$, $B^0_s$--$\bar B^0_s$ mixing, $\overline{\rm BR}(B_s^0 \to \mu^+\mu^-)$, and ${\rm BR}(b \to s \gamma)$. These observables each put an upper bound on $v_{\chi}$ (equivalently $\tan\theta_H$) as a function of $m_3$.  In each case we combine the experimental and GM theoretical uncertainties in quadrature and constrain the GM model prediction for the observable in question to lie within 2$\sigma$ of the experimental central value.  We will refer to these as ``tight'' constraints.

However, in the GM model the $H_3^+$ contributions to $R_b$, $\overline{\rm BR}(B_s^0 \to \mu^+\mu^-)$, and ${\rm BR}(b \to s \gamma)$ worsen the agreement with experiment compared to the SM limit (i.e., compared to taking $v_{\chi} \to 0$ or $m_3 \to \infty$).  As we will see, the SM limit is already 0.8$\sigma$, 1.0$\sigma$, and 1.3$\sigma$ away from the experimental central values of these three observables, respectively.  For this reason, we also consider a second, more conservative approach to constraining the parameter space for these three observables: we require that the GM model prediction lie within 2$\sigma$ of the best-fit value obtainable in the GM model (i.e., the SM limit), again combining the experimental and GM theoretical uncertainties in quadrature. We will refer to these more conservative constraints as ``loose'' constraints.

\begin{figure}
\resizebox{0.5\textwidth}{!}{\includegraphics{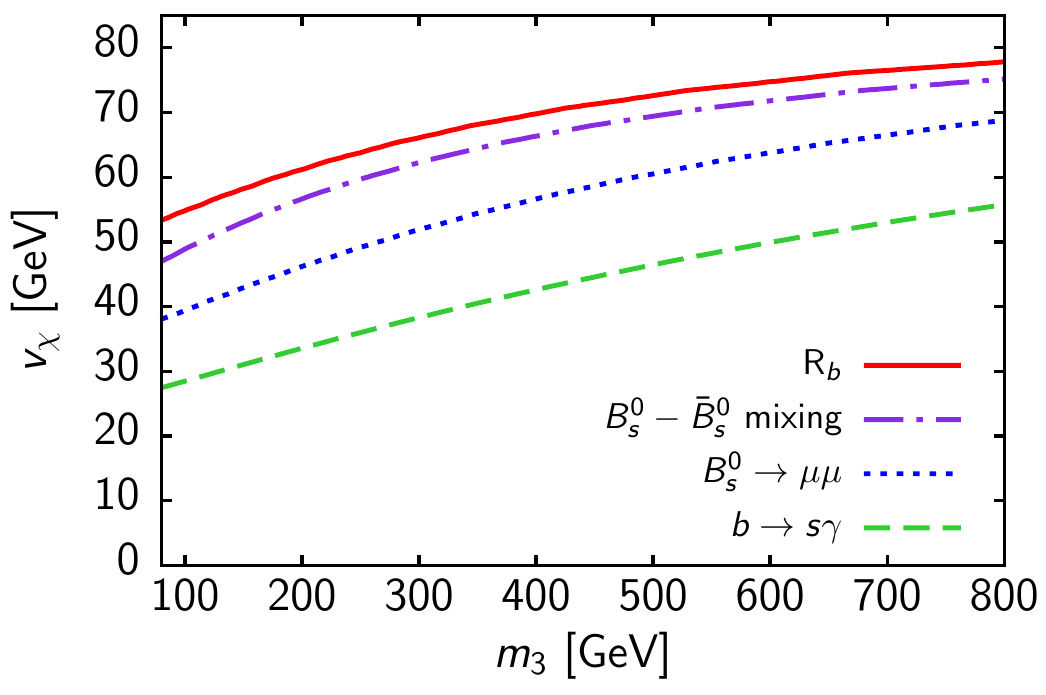}}%
\resizebox{0.5\textwidth}{!}{\includegraphics{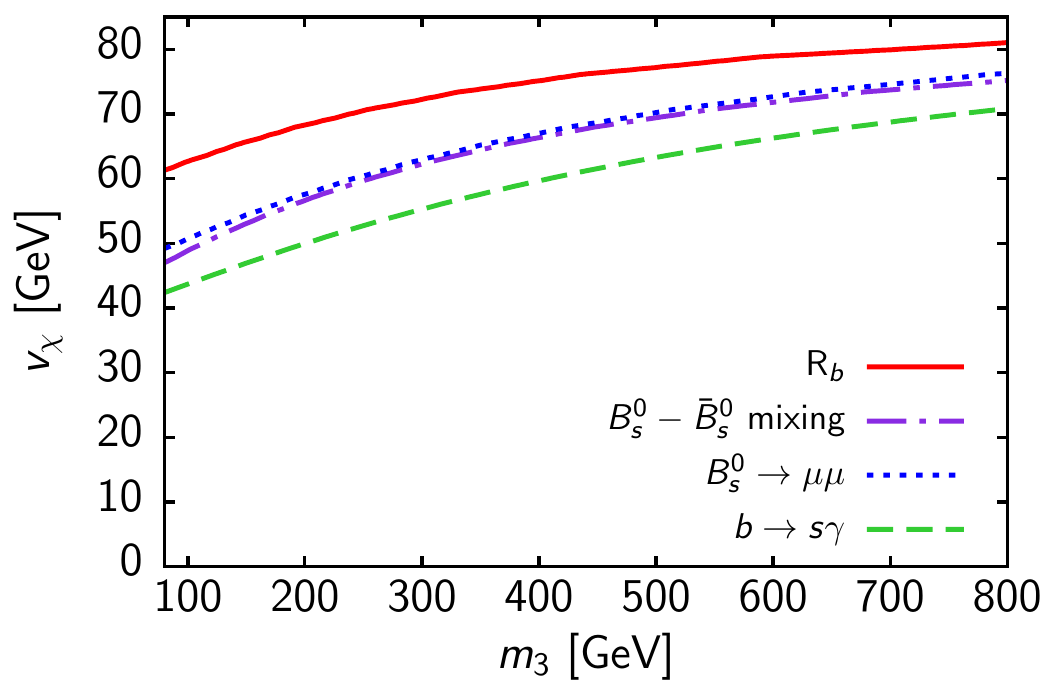}}
\caption{Constraints on $m_3$ and $v_{\chi}$ in the GM model from $R_b$, $B^0_s$--$\bar B^0_s$ mixing, $\overline{\rm BR}(B_s^0 \to \mu^+ \mu^-)$, and ${\rm BR}(b \to s \gamma)$.  The region above each curve is excluded.  Left: ``tight'' constraints requiring that each observable lies within 2$\sigma$ of the experimental central value.  Right: ``loose'' constraints requiring that $R_b$, $\overline{\rm BR}(B_s^0 \to \mu^+ \mu^-)$, and ${\rm BR}(b \to s \gamma)$ lie within 2$\sigma$ of the value at the best-fit point within the GM model.}
\label{fig:expconst}
\end{figure}

These ``loose'' and ``tight'' constraints are respectively shown in the right- and left-hand panels of Fig.~\ref{fig:expconst}.%
\footnote{With the exception of $M_t$, we choose the input parameters for all our numerical results  from the 2014 Review of Particle Physics~\cite{Agashe:2014kda}. For $M_t$, we use the first combination of  Tevatron and LHC measurements of the top quark mass~\cite{ATLAS:2014wva}.
In particular, we set $G_F =1.1663787 \times 10^{-5} \,  {\rm GeV}^{-2} $, $\alpha_{\rm em} = 1/127.94$,  $\alpha_s = 0.1184$, 
$\bar{m}_c(m_c) = 1.275 \, {\rm GeV}$, $\bar{m}_b(m_b) = 4.18 \, {\rm GeV}$, 
$M_Z = 91.1876  \, {\rm GeV} $ and $M_t = 172.9 \, {\rm GeV} $.
In addition, we obtain the dependent parameters $M_W = 79.83~{\rm GeV}$ and $s^2_W=0.2336$ at tree level. We thus edit the input files of SuperIso v3.3 which by default uses inputs from the 2011 Review of Particle Physics~\cite{Nakamura:2010zzi}.
\label{note:parameters}}  
Details on each process follow.


\subsection{$R_b$}
\label{sec:Rb}

The $Z$-pole observable $R_b$, defined as
\begin{equation}
	R_b=\frac{\Gamma(Z\rightarrow b\bar{b})}{\Gamma(Z\rightarrow \rm hadrons)},
\end{equation} 
has been calculated in the SM including two-loop electroweak~\cite{Freitas:2012sy} and three-loop QCD corrections.  The correction to $R_b$ due to one-loop diagrams involving additional Higgs bosons has been calculated in the 2HDM~\cite{Denner:1991ie,Grant:1994ak}. In the Type-I 2HDM, the contribution of the neutral scalars can be neglected~\cite{Grant:1994ak} as it is suppressed by a relative factor of $m_b^2/m_t^2$ compared to the charged Higgs contribution. The results for the Type-I 2HDM can easily be adapted to the GM model~\cite{Haber:1999zh, Chiang:2012cn}.

Following Ref.~\cite{Haber:1999zh}, the one-loop charged Higgs correction to $R_b^{\rm SM}$ can be written as\footnote{The coefficient $-0.7785$ depends on the $Z b \bar{b}$ couplings and the bottom quark mass. Updated values of these quantities have a very 
small effect on the coefficient. For example, using more recent values from Ref.~\cite{Haisch:2011up} the change in the coefficient is 0.1\%.} 
\begin{eqnarray}
	\delta R_b^\text{GM}  &=& -0.7785 \, \delta g^L_{\rm GM} + 0.1409 \, \delta g^R_{\rm GM} \nonumber \\
	&\approx& -0.7785 \, \delta g^L_{\rm GM}      \nonumber \\
	&\simeq& \frac{0.7785}{64\pi^2} \left(\frac{e^3}{s_W^3 c_W}\right) \tan^2\theta_H \,  x_{tW}
		     \left(\frac{x_{t3}}{1-x_{t3}}   +  \frac{x_{t3} \log x_{t3}}{(1  -  x_{t3})^2} \right),
\label{eq:dRb}		     
\end{eqnarray}
 where $x_{tW} = \bar{m}_t^2(\mu_t)/M_W^2$, $x_{t3} = \bar{m}_t^2(\mu_t)/m_3^2$, and we neglect $M_Z$ in the loop calculation.\footnote{Full expressions including the $M_Z$ dependence have been given in Refs.~\cite{Haber:1999zh,Chiang:2012cn}.  Because the constraint from $R_b$ is weaker than the other constraints, we use here only the approximation given in Eq.~(\ref{eq:dRb}).} Here $\bar{m}_t$ is  the 
$\overline{{\rm MS}}$ running mass and is evaluated at $\mu_t = M_Z$.
 The approximation in Eq.~(\ref{eq:dRb}) can be made because $\delta g^R_{\rm GM}$ is suppressed by a factor of $m_b^2/m_t^2$ 
 compared to $\delta g^L_{\rm GM}$~\cite{Haber:1999zh,Chiang:2012cn}. The correction is always negative and interferes destructively with the SM contribution.
 
The measured value of $R_b$ is~\cite{Baak:2014ora}
\begin{equation}
	R_b^{\rm exp} = 0.21629 \pm 0.00066,
\end{equation}
while the SM prediction is $R_b^{\text{SM}}=0.21577\pm0.00011$~\cite{Baak:2014ora}.  Therefore the 2$\sigma$ upper bound relative to the experimental central value yields the ``tight'' constraint $R_b^{\text{GM}}=R_b^{\text{SM}}+\delta R_b^\text{GM} > 0.21495$,
where we have combined the experimental and SM theoretical uncertainties in quadrature.\footnote{Because the coefficients in Eq.~(\ref{eq:dRb}) depend on the SM $R_b$ prediction in a complex way, the $R_b$ observable cannot straightforwardly be calculated using a ratio of the SM and GM constributions (as we will do with the other observables). For this reason, in the $R_b$ case only we take the theory uncertainty on the GM prediction to be the same as that of the SM prediction.}

The SM prediction is 0.8$\sigma$ below the measured value; as the GM correction interferes destructively with the SM contribution, it increases the discrepancy between theory and experiment. As a result, the best agreement with the experimental measurement of $R_b$ in the GM model occurs in the SM limit ($v_\chi\rightarrow 0$, $m_3\rightarrow\infty$).  Requiring that the GM prediction for $R_b$ lie within 2$\sigma$ of the SM value yields a ``loose'' constraint of $R_b^{\text{GM}} > 0.21443$. 

As we see in Fig.~\ref{fig:expconst},  $R_b$ provides the weakest  non-oblique ``tight" and ``loose" constraints. Furthermore, substantial improvement in the $R_b$ experimental measurement is unlikely in the near future, as this would require better Z-pole measurements using  a next-generation $e^+e^-$ collider like the International Linear Collider (ILC) with the GigaZ option~\cite{Baer:2013cma}.


\subsection{$B^0_s$--$\bar{B}^0_s$ mixing}
\label{sec:Bmix}

The effect of charged scalars on $B_d^0$--$\bar{B}_d^0$ mixing in a Type-I 2HDM has been studied~\cite{Athanasiu:1985ie} by adapting results  
of similar processes on $K^0$--$\bar{K}^0$ mixing~\cite{Abbott:1979dt}. These can be extended to the $B^0_s$--$\bar{B}^0_s$ system, which is
more constraining than the $B_d^0$--$\bar{B}_d^0$ system due to lower errors in both the experimental measurement and the SM prediction~\cite{Lenz:2012mb}. 
To leading order, the oscillation frequency of a $B^0_s$ meson in the GM model is determined by the mass 
splitting~\cite{WahabElKaffas:2007xd} 
\begin{equation}
	\Delta m_{B_s}^{\rm GM} = \frac{G_F^2 m_t^2}{24\pi^2} |V_{ts}^*V_{tb}|^2 f_{B_s}^2 B_{B_s} 
	m_{B_s} \eta_b I_{\rm GM}\,.
	\label{eq:DMBs}
\end{equation}
Here $\eta_b$ is a scaling factor, $f_{B_q}$ is the weak decay constant, $B_{B_q}$ is the bag parameter, $m_{B_s}$ is the meson mass, and\footnote{The NLO QCD corrections to the charged Higgs contributions are known~\cite{Urban:1997gw}, but we do not include them here.}
\begin{equation}
	I_{\rm GM} = I_{WW}(x_{tW}) + \tan^2 \theta_H \, I_{WH}(x_{tW},x_{t3},x_{3W})
	+ \tan^4\theta_H  \, I_{HH}(x_{t3}).
\label{eq:IGM}	
\end{equation}
Here $x_{3W}=m_3^2/M_W^2$, $x_{tW}=\bar{m}_t^2(\mu_t)/M_W^2$, and $x_{t3}=\bar{m}_t^2(\mu_t)/m_3^2$.
 We set the top mass renormalization scale $\mu_t=M_t$, where $M_t$ is the top quark pole mass. The Inami-Lim functions~\cite{Inami:1980fz} 
  $I_{WW}$, $I_{HH}$, and $I_{WH}$ are given by \cite{Mahmoudi:2009zx}
\begin{eqnarray}
	I_{WW}(x) &=& 1 + \frac{9}{1-x}-\frac{6}{(1-x)^2}-\frac{6 x^2 \log x}{(1-x)^3}, \nonumber \\
	I_{HH}(x) &=& x\left[\frac{1+x}{(1-x)^2}+\frac{2 x \log x}{(1-x)^3}\right], \nonumber \\
	I_{WH}(x,y,z) &=&  y\left[\frac{(2 z - 8) \log y}{(1-y)^2(1-z)} 
	+ \frac{6 z \log x}{(1-x)^2(1-z)} - \frac{8-2 x}{(1-y)(1-x)}\right].
\end{eqnarray}

Under the assumption that the overall coefficients do not vary substantially due to new scalar contributions, a prediction for $\Delta m_{B_s}^{\rm GM}$ in the GM model may be obtained using the ratio
\begin{equation}
	R_{\Delta m}^{\rm GM} \equiv \frac{\Delta m_{B_s}^{\rm GM}}{\Delta m_{B_s}^{\rm SM}} 
	= 1 + \frac{\tan^2\theta_H \, I_{WH}(x_{tW},x_{t3},x_{3W}) + \tan^4\theta_H  \,   I_{HH}(x_{t3})}{I_{WW}(x_{tW}) }.
\end{equation}
Since $I_{WW}$, $I_{WH}$ and $I_{HH}$ are all positive, the GM model contribution interferes constructively with the SM contribution. Because the theoretical uncertainty on the mass splitting is due almost entirely to uncertainties in the coefficients of $I_{\rm GM}$ in Eq.~(\ref{eq:DMBs}), we scale the SM theoretical uncertainty $\delta \Delta m^{\rm SM}_{B_s}$ by $R^{\rm GM}_{\Delta m}$ to obtain the theoretical uncertainty in the GM model, i.e., $\delta \Delta m^{\rm GM} = R_{\Delta m}^{\rm GM}\cdot \delta\Delta m^{\rm SM}$.

The measured value for the $B_s^0$--$\bar{B}_s^0$ mass difference is given by~\cite{PDG}
\begin{equation}
	\Delta m_{B_s}^{\rm exp} = 17.719 \pm 0.036 {\rm (stat)} \pm 0.023{\rm (syst)}\, {\rm ps^{-1}}.
\end{equation}
The largest uncertainty in the SM prediction comes from the lattice QCD calculation of $f_{B_s} B_{B_s}^{1/2}$.  Using a CKMfitter~\cite{Lenz:2012az,CKMfitter} average of several lattice results based on separate extractions of $f_{B_s}$ and $B_{B_s}$, Ref.~\cite{Nierste:2012qp} obtains the SM prediction
$\Delta m_{B_s}^{\rm SM} = 17.3 \pm 1.5 \ {\rm ps^{-1}}$.  
However, a preliminary lattice calculation of the product $f_{B_s} B_{B_s}^{1/2}$ from the Fermilab/MILC collaboration~\cite{Bouchard:2011xj} yields a larger uncertainty and considerably different central value, leading to $\Delta m_{B_s}^{\rm SM} = 21.7 \pm 2.6 \ {\rm ps^{-1}}$~\cite{Nierste:2012qp}.  For our numerical results we use the CKMfitter central value but take the more conservative uncertainty as advocated in Ref.~\cite{Lenz:2011ti},
\begin{equation}
	\Delta m_{B_s}^{\rm SM} = 17.3 \pm 2.6 \ {\rm ps^{-1}}.
\end{equation}

The above results can be translated into an experimental measurement of $R_{\Delta m}^{\rm exp} \equiv \Delta m_{B_s}^{\rm exp}/\Delta m_{B_s}^{\rm SM}$ and combined experimental and SM theoretical uncertainty of
\begin{equation}
	R^{\rm exp}_{\Delta m} = 1.02 \pm 0.15.
\end{equation}
The SM prediction $R_{\Delta m}^{\rm SM} =1$ is thus only 0.13$\sigma$ below the measured value. 

In the case of $B^0_s$--$\bar{B}^0_s$ mixing, the charged Higgs contributions in the GM model increase the predicted value of $R_{\Delta m}$, so that the best-fit value and the experimental central value are the same.  Thus our ``tight'' and ``loose'' constraints from this observable are the same.
The 2$\sigma$ constraint is $R_{\Delta m}^{\rm GM}\leq 1.46$, where we have combined the experimental and GM theoretical uncertainties in quadrature.  The resulting constraint on the $m_3$--$v_{\chi}$ plane is shown in the left and right panels of Fig.~\ref{fig:expconst}. It is slightly more constraining than the bounds from $R_b$, and is about the same as the ``loose'' bound from $B_s^0 \rightarrow \mu^+\mu^-$.\footnote{If we were to use the less-conservative prediction of $\Delta m_{B_s}^{\rm SM} = 17.3 \pm 1.5~{\rm ps}^{-1}$, the uncertainty on $R^{\rm exp}_{\Delta m}$ becomes $0.089$ and the bound would tighten to match the ``tight'' bound from $B_s^0 \rightarrow \mu^+\mu^-$ in the left-hand panel. If we were instead to use the central value $\Delta m_{B_s}^{\rm SM} = 21.7 \pm 2.6~{\rm ps}^{-1}$, the best-fit reference point would become the SM prediction, $R^{\rm exp}_{\Delta m} = 0.817 \pm 0.098$, and the ``loose'' and ``tight'' bounds would each be slightly stronger than the corresponding bounds from $b\rightarrow s\gamma$.  This variability illustrates the very large remaining theoretical uncertainty in this observable.} In both cases the bound is weaker than that from $b\rightarrow s\gamma$.  An improvement in the constraint from $B_s^0$--$\bar B_s^0$ mixing relies on an improved lattice determination of $f_{B_s} B^{1/2}_{B_s}$.

\subsection{$B_s^0 \rightarrow \mu^+\mu^-$}
\label{sec:Bmu}

A full leading-order computation of the average time-integrated branching ratio $\overline{\rm BR}(B_s^0 \to \mu^+ \mu^-)$ in the Aligned 2HDM~\cite{Pich:2009sp} was recently performed in Ref.~\cite{Li:2014fea}.  The calculation can be easily specialized to the Type-I 2HDM and hence to the GM model; the result is conveniently expressed in terms of a ratio to the SM prediction,
\begin{equation}
	\overline{R}_{s\mu}^{\rm GM} = \frac{\overline{\rm BR}(B_s^0\rightarrow \mu^+\mu^-)_{\rm GM}}
	{\overline{\rm BR}(B_s^0\rightarrow \mu^+\mu^-)_{\rm SM}} 
	\simeq \left|\frac{C_{10}^{\rm GM}}{C_{10}^{\rm SM}}\right|^2,
	\label{eq:Rsmu}
\end{equation}
where the Wilson coefficients $C_{10}^{\rm SM}$ and $C_{10}^{\rm GM}$ are given by~\cite{Li:2014fea}
\begin{eqnarray}
	C_{10}^{\rm SM} &=& -0.9380 \left[\frac{M_t}{173.1 {\rm GeV}}\right]^{1.53}
	\left[\frac{\alpha_s(M_Z)}{0.1184}\right]^{-0.09}, \nonumber \\
	C_{10}^{\rm GM} &=& C_{10}^{\rm SM} + \tan^2\theta_H \frac{x_{tW}}{8}
	\left[ \frac{x_{t3}}{1-x_{t3}} + \frac{x_{t3} \log x_{t3}}{(1-x_{t3})^2} \right],
	\label{eq:C10GM}
\end{eqnarray}
with $x_{tW}=\bar{m}_t^2(\mu_t)/M_W^2$ and $x_{t3} \equiv m_t^2(\mu_t)/m_3^2$ as before.\footnote{Note that for $x_{t3} \to 1$, the expression in square brackets can be expanded in powers of $\delta \equiv x_{t3} - 1$ and reads
\begin{equation}
	\left[ \frac{x_{t3}}{1-x_{t3}} + \frac{x_{t3} \log x_{t3}}{(1-x_{t3})^2} \right] \simeq - \frac{1}{2} - \frac{\delta}{6} + \mathcal{O}(\delta^2).
\end{equation}} Here $\mu_t=M_t$ is the top quark pole mass and $\bar m_t$ is the $\overline{\rm MS}$ running mass.  The theoretical uncertainty on the resulting GM branching ratio is taken to be $\delta \overline{\rm BR}(B_s^0 \to \mu^+ \mu^-)_{\rm GM} = \overline{R}_{s\mu}^{\rm GM} \cdot \delta\overline{\rm BR}(B_s^0\rightarrow \mu^+\mu^-)_{\rm SM}$.

The approximation made in Eq.~(\ref{eq:Rsmu}) is to neglect contributions from the Wilson coefficients $C_S$ and $C_P$, which arise from scalar and pseudoscalar penguins and box diagrams and are suppressed by an extra factor of $m_b^2/m_t^2$ compared to $C_{10}$.  The expression for $C_{10}^{\rm SM}$ includes next-to-leading order (NLO) electroweak and QED corrections, as well as NLO and next-to-next-to-leading order (NNLO) QCD corrections.
We note that the expression for the $H_3^+$ contribution to $C_{10}^{\rm GM}$ has the same dependence on $m_3$ and $\tan\theta_H$ as the expression for the correction to $R_b$ in the $M_Z \to 0$ limit given in Eq.~(\ref{eq:dRb}).  This is because the charged Higgs contribution to $C_{10}^{\rm GM}$ comes from the same $Z$ penguin diagrams as in $R_b$, but with a generation-changing $H^+ \bar t_R s_L$ vertex in place of the generation-conserving $H^+ \bar t_R b_L$ vertex and $p^2_Z = M_{B_s}^2 \simeq 0$.

The current experimental measurement of $\overline{\rm BR}(B_s^0 \to \mu^+ \mu^-)$ from a combination of CMS and LHCb results is~\cite{CMSandLHCbCollaborations:2013pla}
\begin{equation}
	\overline{\rm BR}(B_s^0 \to \mu^+\mu^-)_{\rm exp} = (2.9 \pm 0.7) \times 10^{-9}.
\end{equation}
The SM prediction is given in Ref.~\cite{Li:2014fea} as
\begin{equation}
	\overline{\rm BR}(B_s^0 \to \mu^+\mu^-)_{\rm SM} = (3.67 \pm 0.25)\times 10^{-9}.
\end{equation}
This number differs slightly from the result in Ref.~\cite{Bobeth:2013uxa}, upon which it is based, due to the use of a slightly different central value and more conservative uncertainty on the top quark mass.  These yield an experimental measurement of $\overline{R}_{s\mu}$ and combined experimental and SM theoretical uncertainty of
\begin{equation}
	\overline{R}_{s \mu}^{\rm exp} = 0.79 \pm 0.20.
\end{equation}
In particular, the SM prediction, $\overline{R}_{s \mu}^{\rm SM} = 1$, is 1.0$\sigma$ above the current experimental value.

The 2$\sigma$ constraint on the GM model relative to the experimental central value yields a bound of $\overline{R}_{s \mu} \leq 1.21$, where we have combined the experimental and GM model theoretical uncertainties in quadrature. This ``tight" constraint is shown in the left-hand panel of Fig.~\ref{fig:expconst}; it is stronger than the corresponding constraints from $R_b$ and $B^0_s$--$\bar{B}^0_s$ mixing, as previously discussed, but remains weaker than the ``tight" $b\rightarrow s\gamma$ constraint.

However, the GM model contribution to $C_{10}$ in Eq.~(\ref{eq:C10GM}) is always negative, leading to constructive interference with the SM contribution and increasing the prediction for $\overline{R}_{s \mu}$ compared to its value in the SM.  As the SM value is already larger than the experimental value ($\overline{R}_{s \mu}^{\rm exp} <1$), the best agreement with the experimental measurement of $\overline{\rm BR}(B_s^0 \to \mu^+ \mu^-)$ in the GM model occurs in the limit $v_{\chi} \to 0$ or $m_3 \to \infty$ (i.e., the SM limit). The best-fit 2$\sigma$ bound taken relative to the SM prediction yields a ``loose" constraint of $\overline{R}_{s \mu} \leq 1.43$, which is shown in the right-hand panel of Fig.~\ref{fig:expconst}.  The ``loose'' constraint from $\overline{\rm BR}(B_s^0 \to \mu^+ \mu^-)$ is also weaker than that from ${\rm BR}(b \to s \gamma)$.

The current uncertainty on $\overline{R}_{s\mu}$ is dominated by the experimental statistical uncertainty.  This has the potential to be significantly reduced in the near future as more data is collected at the LHC.  In particular, the upgraded LHCb experiment is expected to measure $\overline{\rm BR}(B_s^0 \to \mu^+ \mu^-)$ with an ultimate experimental uncertainty of better than 10\% with 50 fb$^{-1}$ of data~\cite{Bediaga:2012py}, which corresponds to about ten years of LHC running.  Assuming an experimental rate consistent with the SM prediction and no change in the theoretical uncertainty, this would correspond to a combined uncertainty on $\overline{R}^{\rm exp}_{s \mu}$ of 0.12.  This measurement thus has the potential to become the most stringent constraint on the GM model parameter space in the near future.

\subsection{$b \to s \gamma$}
\label{sec:bsgam}

The $b\rightarrow s \gamma$ branching ratio has been measured at several different experiments, including CLEO, BaBar, Belle, and ALEPH. The current experimental average from the Heavy Flavour Averaging Group is~\cite{Asner:2010qj,PDG}\footnote{The most recent measurement from BaBar, which has not yet been incorporated into this average, reads ${\rm BR}(\bar B \to X_s \gamma) = (3.29\pm0.19\pm0.48)\times 10^{-4}$~\cite{Lees:2012ufa}.}
\begin{equation}
	{\rm BR}(\bar{B} \to X_s\gamma)_{\rm exp} = (3.55\pm 0.24 \pm 0.09) \times 10^{-4},
\end{equation}
for a photon energy $E_{\gamma} > 1.6$~GeV.

${\rm BR}(b \to s \gamma)$ is known up to NNLO in QCD in the SM~\cite{Misiak:2006zs,Becher:2006pu}.\footnote{This calculation is an estimate insofar as charm-mass-dependent contributions have been incorporated using an interpolation in $m_c$, resulting in a contribution to the theory uncertainty from the interpolation ambiguity.}  The two current SM predictions are ${\rm BR}(\bar B \to X_s \gamma)_{\rm SM} = (3.15 \pm 0.23) \times 10^{-4}$~\cite{Misiak:2006zs} and ${\rm BR}(\bar B \to X_s \gamma)_{\rm SM} = (2.98 \pm 0.26) \times 10^{-4}$~\cite{Becher:2006pu}. These predictions differ due to different approaches to handling higher-order contributions to the photon energy cutoff corrections; however, their difference is within the $\pm 3\%$ theoretical uncertainty due to uncalculated higher orders~\cite{Misiak:2006zs}.

The charged Higgs contributions in the Type-I 2HDM, first calculated in Ref.~\cite{Barger:1989fj}, are themselves now known up to NLO in QCD~\cite{Ciuchini:1997xe}.  Because ${\rm BR}(b \to s \gamma)$ will provide the most stringent constraint on the GM model parameter space, we will use the full implementation of the SM and 2HDM contributions in the public code SuperIso~v3.3~\cite{SuperIso}, which is based on the calculations in Refs.~\cite{Misiak:2006zs,Misiak:2006ab}.  SuperIso calls the code 2HDMC~v1.6.4~\cite{2HDMC} for spectrum calculations within the Type-I 2HDM.

In the limit $v_{\chi} \to 0$ or $m_3 \to \infty$, the calculation of ${\rm BR}(\bar B \to X_s \gamma)$ by SuperIso~v3.3, using the input parameters given in
footnote~\ref{note:parameters}, yields a prediction
\begin{equation}
	{\rm BR}(\bar B \to X_s \gamma)_{\rm SM \ limit} = 3.11 \times 10^{-4}.
\end{equation}
The difference compared to the SM predictions quoted above is primarily due to differences in the input parameters, particularly $m_b$ and $m_c$~\cite{Nazila}.  However, the difference is still within the theoretical uncertainty due to parametric uncertainties of $\pm 3\%$~\cite{Misiak:2006zs}.  We take the total theoretical uncertainty on this SM prediction to be $\pm 0.23 \times 10^{-4}$ from Ref.~\cite{Misiak:2006zs}.  Combining this in quadrature with the experimental uncertainty yields a total uncertainty of $\pm 0.34 \times 10^{-4}$.  In particular, the value of ${\rm BR}(\bar B \to X_s \gamma)$ in the SM limit is 1.3$\sigma$ below the experimental value.

The charged Higgs contribution to ${\rm BR}(\bar B \to X_s \gamma)$ in the GM model interferes destructively with the SM contribution, leading to a smaller predicted value for ${\rm BR}(\bar B \to X_s \gamma)$ than in the SM.  Because the SM prediction is already below the experimental central value, the best agreement with the experimental measurement in the GM model occurs in the limit $v_{\chi} \to 0$ or $m_3 \to \infty$ (i.e., the SM limit). Since even the SM limit yields a prediction that is only 0.7$\sigma$ from the experimental bound, the 2$\sigma$ experimental constraint on the GM $m_3$--$v_{\chi}$ plane is quite strong, as can be seen in the left panel of Fig.~\ref{fig:expconst}. This bound corresponds to ${\rm BR}(\bar B \to X_s \gamma) > 2.88 \times 10^{-4}$ (``tight'' constraint), where we have combined the experimental and GM theoretical uncertainties in quadrature; again, here we estimate the GM theory uncertainty to be that of the SM prediction scaled by a ratio of the GM and SM predictions. In comparison, the $2\sigma$ constraint with respect to the best-fit point, the SM limit, yields ${\rm BR}(\bar B \to X_s \gamma) > 2.48 \times 10^{-4}$ (``loose'' constraint).  This is shown in the right panel of Fig.~\ref{fig:expconst} together with the ``loose'' constraints from the other observables.  In either case, ${\rm BR}(\bar B \to X_s \gamma)$ is the strongest constraint on these parameters.

Because of the large theoretical uncertainty on ${\rm BR}(\bar B \to X_s \gamma)$ and the sensitivity of the resulting constraint to the particular choice of input parameters and the handling of partial higher-order corrections, we consider it safer to take the more conservative approach and apply the ``loose'' constraint from ${\rm BR}(\bar B \to X_s \gamma)$ as our primary constraint on the $m_3$--$v_{\chi}$ plane.  We will nevertheless also show the effect of applying the ``tight'' $b \to s \gamma$ constraint in our numerical scans.

The current theoretical and experimental uncertainties on ${\rm BR}(\bar B \to X_s \gamma)$ are comparable in size.  The experimental uncertainty is expected to be reduced with measurements at the super $B$ factory experiment Belle~II currently under construction at KEK.  A conservative treatment of systematics yields an estimated future experimental precision on ${\rm BR}(\bar B \to X_s \gamma)$ of 7\% (i.e., about $\pm 0.21 \times 10^{-4}$) with 5~ab$^{-1}$ of data, or 6\% (i.e., about $\pm 0.18 \times 10^{-4}$) with 50~ab$^{-1}$ of data~\cite{Aushev:2010bq}.  With the current theoretical uncertainties, these would reduce the combined uncertainty only to about $\pm 0.31 \times 10^{-4}$ or $\pm 0.29 \times 10^{-4}$, respectively.  A more significant improvement in the constraining power of ${\rm BR}(\bar B \to X_s \gamma)$ would require a simultaneous reduction in the theoretical uncertainty.

\section{Numerical results}
\label{sec:numerical}

We now illustrate the effects of the indirect experimental constraints from ${\rm BR}(b \to s \gamma)$ (computed using SuperIso v3.3~\cite{SuperIso}, which calls 2HDMC~v1.6.4~\cite{2HDMC}) and the $S$ parameter on the parameter space of the GM model.  We scan over the full range of GM model parameters allowed after imposing the theoretical requirements of perturbative unitarity, bounded-from-belowness of the potential, and the absence of alternative custodial-symmetry--breaking minima~\cite{HKL}.  We require that either $h$ or $H$ has mass 125~GeV and set the SM Higgs vev $v$ using $G_F$.  We take $\mu_3^2 \leq (1200~{\rm GeV})^2$, which fully populates the mass ranges shown in Figs.~\ref{fig:mvchi}--\ref{fig:kfkVmass} below.  In Fig.~\ref{fig:kfkVmass} we will include additional points generated by a dedicated scan with $\mu_3^2 \leq (200~{\rm GeV})^2$ in order to better populate the low-mass region.  In all cases, we show the effects of the following constraints:
\begin{itemize}
\item The prediction for $S$ yields $\chi^2 \leq 4$ after marginalizing over the $T$ parameter.  Points eliminated by this constraint are shown by red (medium gray) $+$ shapes.
\item The prediction for ${\rm BR}(b \to s \gamma)$ lies within $2\sigma$ of the model point that gives the best agreement with the experimental measurement (``loose'' constraint).  We combine theoretical and experimental uncertainties in quadrature.  Points eliminated by this constraint are shown by light green (light gray) $\times$ shapes.
\item The prediction for ${\rm BR}(b \to s \gamma)$ lies within $2\sigma$ of the experimental measurement (``tight'' constraint).  We combine theoretical and experimental uncertainties in quadrature.  Points eliminated by this constraint are shown by dark green (dark gray) $\times$ shapes.
\end{itemize}
Points depicted in black are allowed by all constraints.

We start by showing the effect of the $b \to s \gamma$ measurement on the $m_3$--$v_{\chi}$ plane in the left panel of Fig.~\ref{fig:mvchi}.  The prediction for ${\rm BR}(b \to s \gamma)$ in the GM model depends only on these two parameters.  We see that, due to its interplay with the decoupling effect of falling $v_{\chi}$ with increasing triplet masses~\cite{HKL}, the ``loose'' $b \to s \gamma$ constraint eliminates all model points with $v_{\chi} \gtrsim 65$~GeV and the ``tight'' $b \to s \gamma$ constraint eliminates all model points with $v_{\chi} \gtrsim 54$~GeV.\footnote{This constraint is considerably more stringent than the upper bound on $v_{\chi}$ obtained in the spirit of Ref.~\cite{Barger:1989fj} by requiring $\cot \theta_H > 0.3$ to avoid parameter regions in which the top quark Yukawa coupling becomes too large ($\cot\theta_H$ plays the same role as is played by $\tan\beta$ in the Type-I 2HDM); this requirement yields $v_{\chi} < 83$~GeV.}

\begin{figure}
\resizebox{0.5\textwidth}{!}{\includegraphics{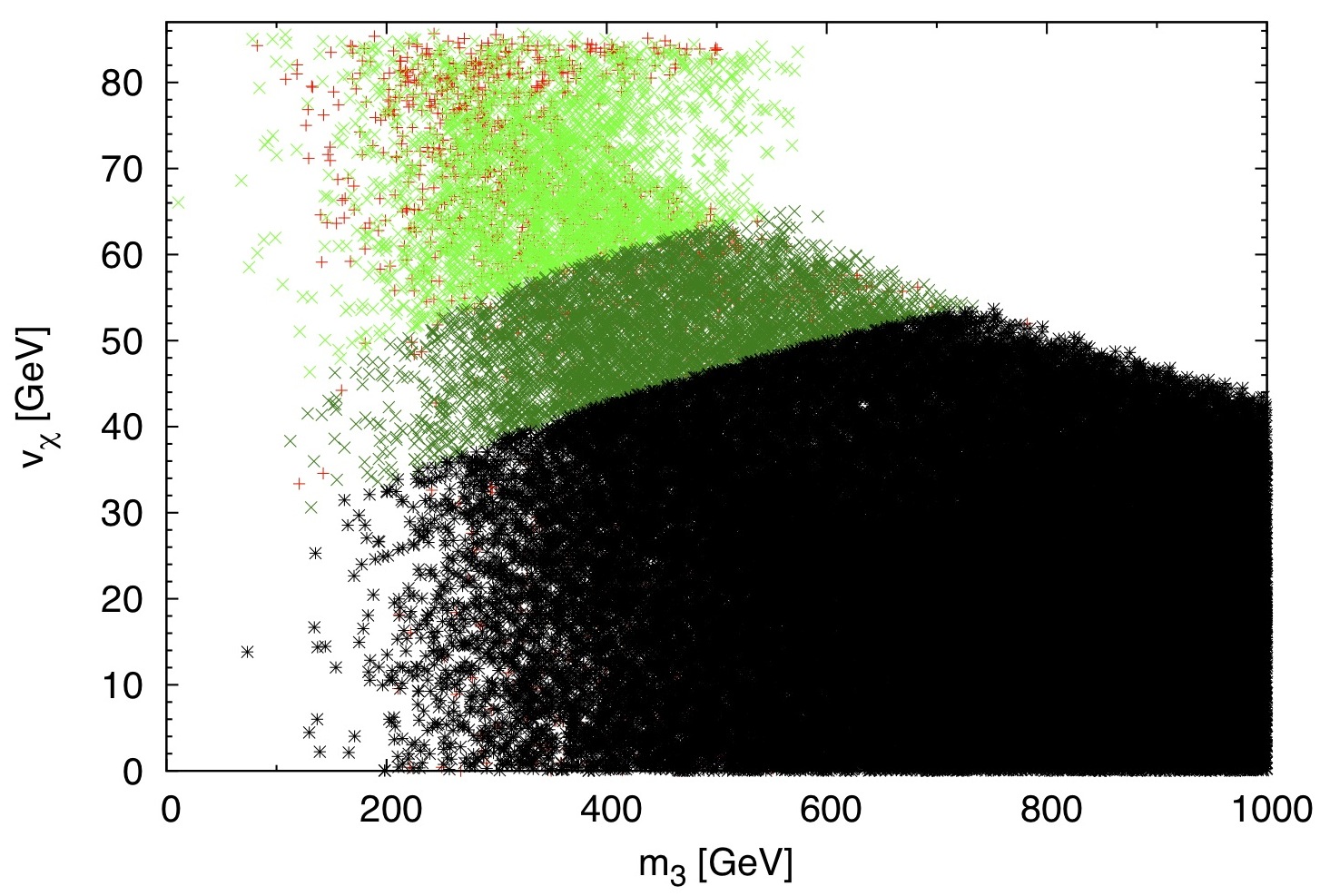}}%
\resizebox{0.5\textwidth}{!}{\includegraphics{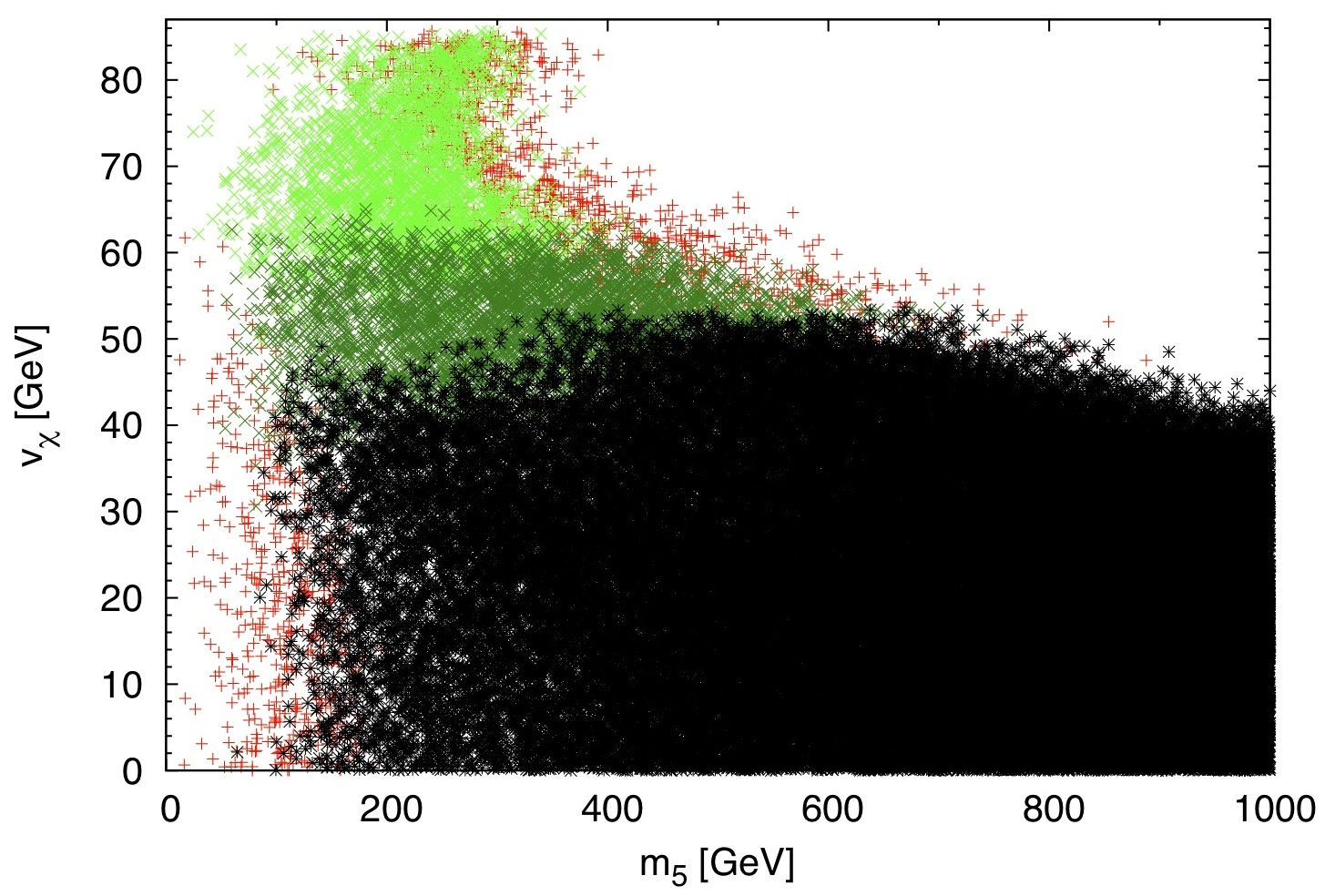}}
\caption{Effect of the experimental constraints on ${\rm BR}(b \to s \gamma)$ and the $S$ parameter on $v_{\chi}$, as a function of $m_3$ (left) and $m_5$ (right).  The black points are allowed.  The red (medium gray) $+$-shaped points are eliminated by the $S$ parameter constraint.  The light green (light gray) $\times$-shaped points are eliminated by the ``loose'' $b \to s \gamma$ constraint, in which we require that ${\rm BR}(b \to s \gamma)$ is within $2\sigma$ of the best-fit point in the GM model.  The dark green (dark gray) $\times$-shaped points would be eliminated by the ``tight'' $b \to s \gamma$ constraint, in which we require that ${\rm BR}(b \to s \gamma)$ lies within $2\sigma$ of the experimental central value.}
\label{fig:mvchi}
\end{figure}

This is reflected in the right panel of Fig.~\ref{fig:mvchi}, where we plot $v_{\chi}$ as a function of $m_5$.  Because $m_5 \neq m_3$ in general, values of $v_{\chi}$ up to the limit of $\sim 65$~GeV are allowed even for $m_5$ masses as low as 100~GeV under the ``loose'' $b \to s \gamma$ constraint.  

This indirect constraint on $v_{\chi}$ as a function of $m_5$ is especially interesting in light of the recent recasting~\cite{Chiang:2014bia} of an ATLAS measurement~\cite{ATLAS-CONF-2014-013} of the like-sign $WWjj$ cross section in 8~TeV data in the context of the GM model.  The like-sign $WWjj$ cross section receives contributions especially from the $s$-channel production of $H_5^{++}$ in $W^+W^+$ fusion, followed by decays back to $W^+W^+$.  The analysis of Ref.~\cite{Chiang:2014bia} excludes a triangular region of parameter space in the $m_5$--$v_{\chi}$ plane extending from $m_5 \simeq 120$~GeV to 610~GeV at $v_{\chi} = 65$~GeV, down to $v_{\chi} \simeq 33$~GeV at $m_5 \simeq 200$~GeV.

In the right panel of Fig.~\ref{fig:mvchi} we also see the effect of the $S$ parameter constraint, which eliminates a few model points at very low $m_5$, as well as moderate to high $m_5$ and high $v_{\chi}$.  This is illustrated in more detail in Fig.~\ref{fig:m5}.  In particular, points with very low values of $m_5$ tend to have a large $m_5$--$m_3$ splitting, which leads to large positive values of the $S$ parameter.

\begin{figure}
\resizebox{0.5\textwidth}{!}{\includegraphics{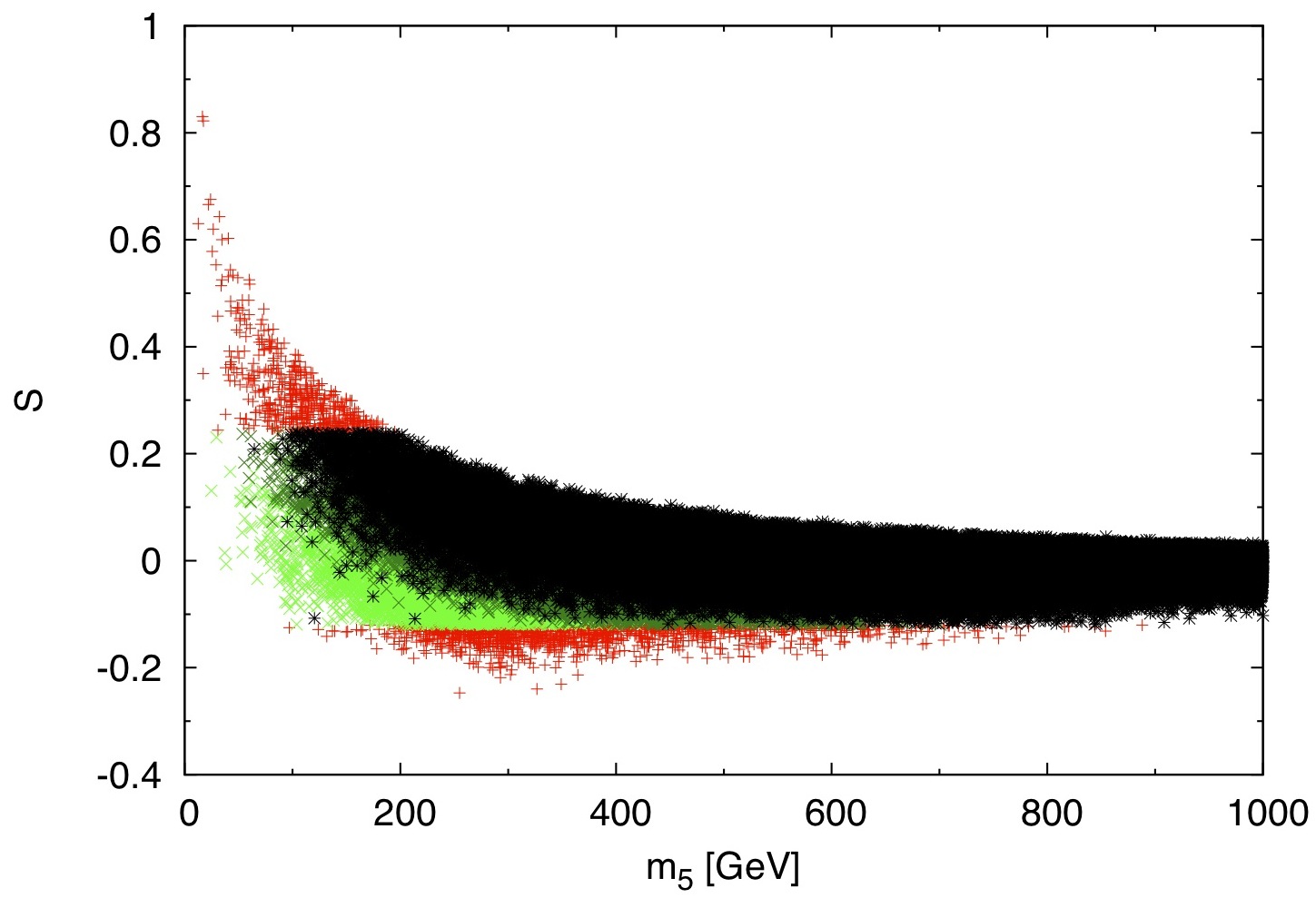}}%
\resizebox{0.5\textwidth}{!}{\includegraphics{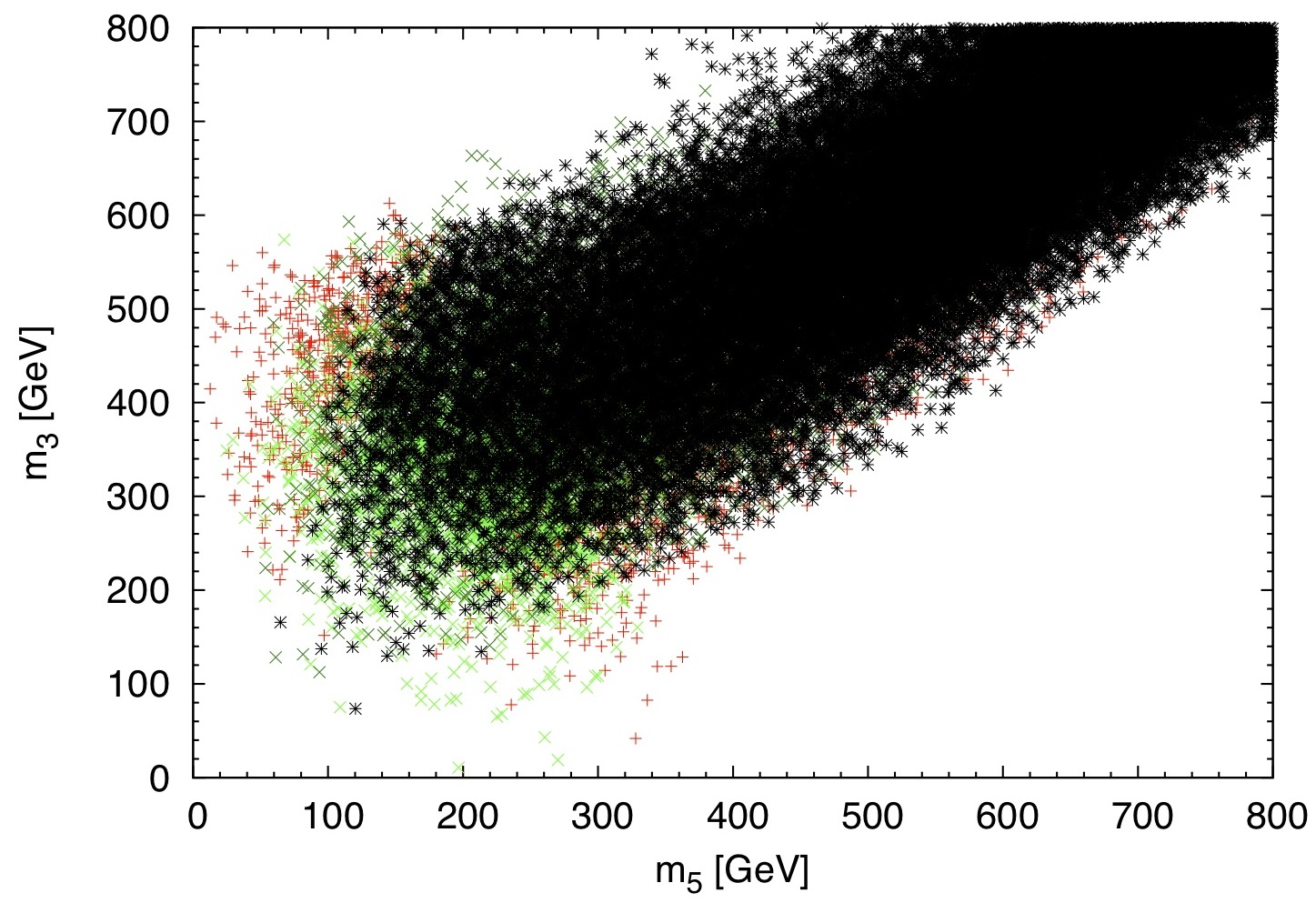}}
\caption{Effect of the experimental constraints on ${\rm BR}(b \to s \gamma)$ and the $S$ parameter as a function of $m_5$.  The color codes are the same as in Fig.~\ref{fig:mvchi}.}
\label{fig:m5}
\end{figure}

Finally we show the effect of the constraints from ${\rm BR}(b \to s \gamma)$ and the $S$ parameter on the allowed ranges of the couplings of the 125~GeV Higgs boson $h$ to $W$ and $Z$ boson pairs and to fermion pairs.  We parameterize these couplings in terms of scaling factors $\kappa_V$ and $\kappa_f$~\cite{LHCHiggsCrossSectionWorkingGroup:2012nn}, which represent the $hVV$ ($V = W,Z$) and $h f\bar f$ couplings, respectively, normalized to their SM values.  In the GM model, these couplings are given in terms of the triplet vev $v_{\chi}$ and the custodial singlet mixing angle $\alpha$ by
\begin{equation}
	\kappa_V = \cos\alpha \frac{v_{\phi}}{v} - \frac{8}{\sqrt{3}} \sin\alpha \frac{v_{\chi}}{v}, 
		\qquad \qquad
	\kappa_f = \cos\alpha \frac{v}{v_{\phi}},
\end{equation}
where $v^2 = v_{\phi}^2 + 8 v_{\chi}^2 \simeq (246~{\rm GeV})^2$ corresponds to the SM Higgs vev.\footnote{For a small number of points in our scan, the 125~GeV Higgs boson is $H$, and the lighter custodial singlet $h$ has a mass below 125~GeV.  In these cases, we plot the coupling scaling factors $\kappa_V$ and $\kappa_f$ that represent the $HVV$ ($V = W,Z$) and $H f\bar f$ couplings, respectively, normalized to their SM values.  These couplings are given in this case by
\begin{equation}
	\kappa_V = \sin\alpha \frac{v_{\phi}}{v} + \frac{8}{\sqrt{3}} \cos\alpha \frac{v_{\chi}}{v}, 
		\qquad \qquad
	\kappa_f = \sin\alpha \frac{v}{v_{\phi}}.
\end{equation}}

One of the most interesting features of the GM model is the possibility that $\kappa_V > 1$, which is not possible at tree level in extended Higgs sectors that contain only SU(2)$_L$ doublets and/or singlets.  In particular, for maximal $v_{\chi}$ and $|\sin\alpha| \sim 1$ (corresponding to $h$ being entirely composed of triplet), $\kappa_V$ can be as large as 1.6.  This maximal value is reduced to $\kappa_V \lesssim 1.36$ by the upper bound on $v_{\chi}$ imposed by the ``loose'' $b \to s \gamma$ constraint, as shown in the left panel of Fig.~\ref{fig:kappas}.  It would be reduced even further to $\kappa_V \lesssim 1.28$ by the ``tight'' $b \to s \gamma$ constraint.

\begin{figure}
\resizebox{0.5\textwidth}{!}{\includegraphics{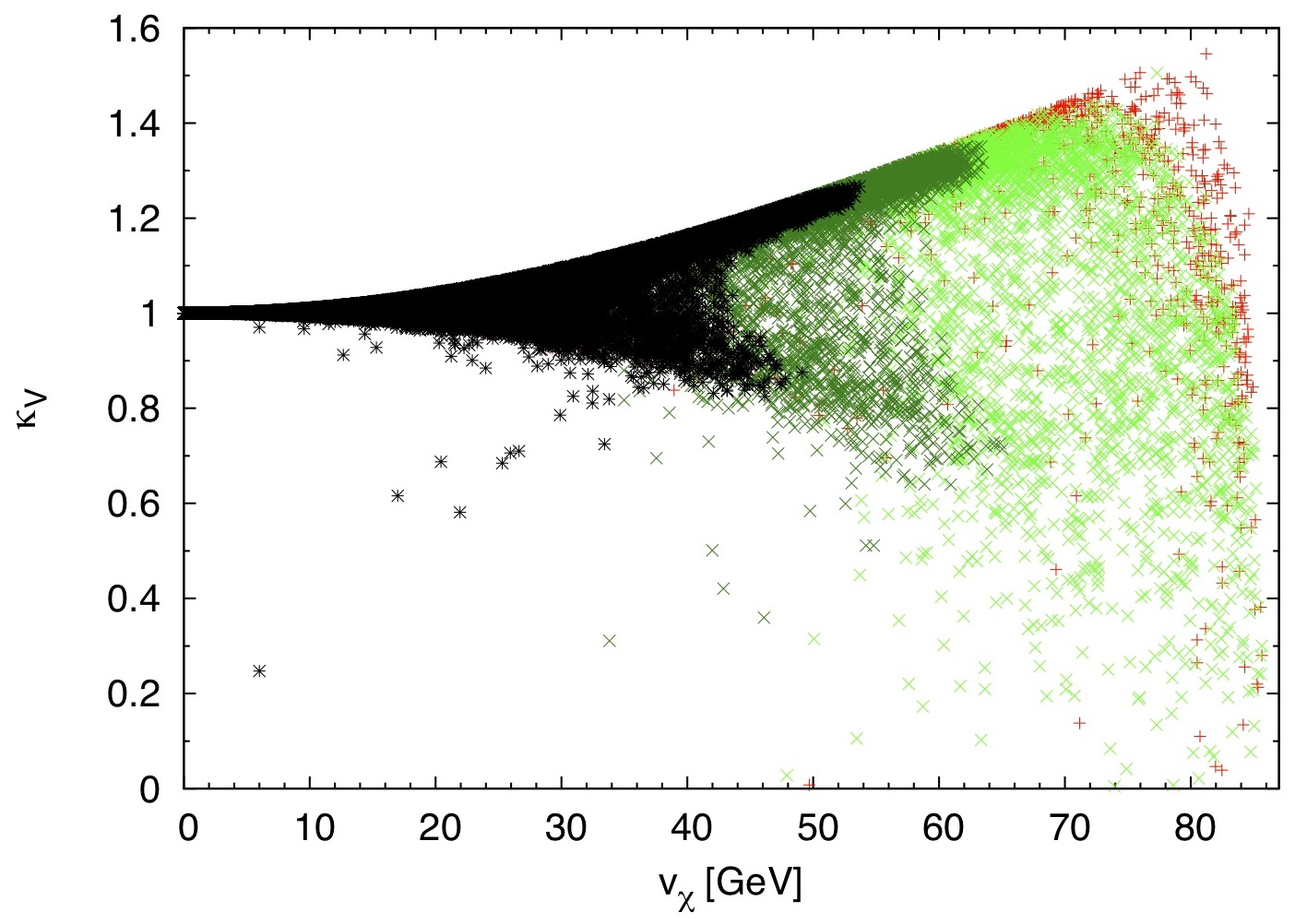}}%
\resizebox{0.5\textwidth}{!}{\includegraphics{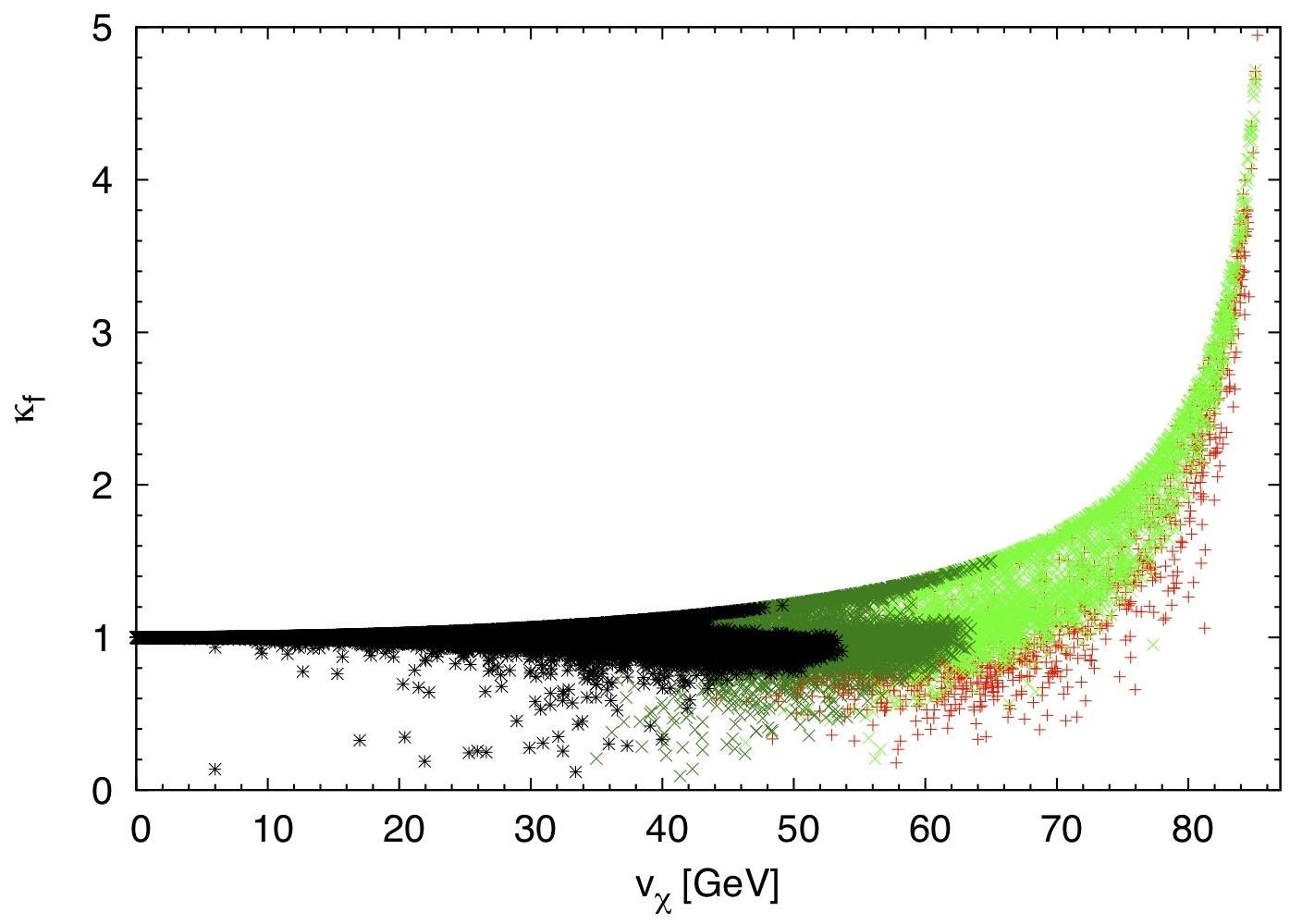}}
\caption{Effect of the experimental constraints on ${\rm BR}(b \to s \gamma)$ and the $S$ parameter on the couplings of the 125~GeV Higgs boson $h$ to $W$ and $Z$ boson pairs (left) and fermion pairs (right), shown as a function of $v_{\chi}$.  $\kappa_V$ ($\kappa_f$) is defined as the coupling of $h$ to $VV$ ($f \bar f$) normalized to the corresponding SM Higgs boson coupling.  The color codes are the same as in Fig.~\ref{fig:mvchi}.}
\label{fig:kappas}
\end{figure}

Similarly, $\kappa_f$ can be significantly enhanced in the GM model if $v_{\phi}$ is small.  By limiting the maximum size of $v_{\chi}$, the ``loose'' $b \to s \gamma$ constraint puts a lower bound on $v_{\phi}$, and thereby imposes an upper bound on $\kappa_f$ of $\kappa_f \lesssim 1.49$, as shown in the right panel of Fig.~\ref{fig:kappas}.  The ``tight'' constraint on $b \to s \gamma$ would reduce this further to $\kappa_f \lesssim 1.20$.

The $S$ parameter measurement does not further constrain the allowed ranges of either $\kappa_V$ or $\kappa_f$ in a significant way once either of the $b \to s \gamma$ constraints has been applied.

Finally we show the allowed range of correlations between $\kappa_V$ and $\kappa_f$ in Fig.~\ref{fig:kvkfcor}.  We note in particular that the GM model can accommodate \emph{simultaneous} enhancements of both $\kappa_V$ and $\kappa_f$.  Such enhancements are constrained by the $b \to s \gamma$ measurement to lie below $\kappa_V \simeq \kappa_f \simeq 1.18$ (``loose'' constraint).  The ``tight'' $b \to s\gamma$ constraint would reduce this to about 1.09.  This is interesting primarily because Higgs coupling fits from LHC data suffer from a flat direction~\cite{Zeppenfeld:2000td} if unobserved decay modes are allowed, corresponding to a simultaneous increase in the unobserved decay branching ratio and in all the Higgs couplings to SM particles.  This flat direction can be cut off by imposing additional theory assumptions, such as the absence of new, unobserved Higgs decay modes~\cite{Zeppenfeld:2000td} or the imposition of $\kappa_V \leq 1$ valid when the Higgs sector contains only isospin doublets and/or singlets~\cite{Duhrssen:2004cv}.  The GM model provides a concrete example of a model that violates the second assumption while being consistent with other experimental constraints.  The flat direction could also be tamed by constraining the total Higgs width through measurements of off-shell $gg \to h^* \to ZZ$~\cite{Caola:2013yja,CMS:2014ala}; the interpretation of this measurement in terms of a Higgs width constraint, however, is itself model-dependent~\cite{Englert:2014aca} and it is not yet clear what effect the presence of additional Higgs states will have.

\begin{figure}
\resizebox{0.5\textwidth}{!}{\includegraphics{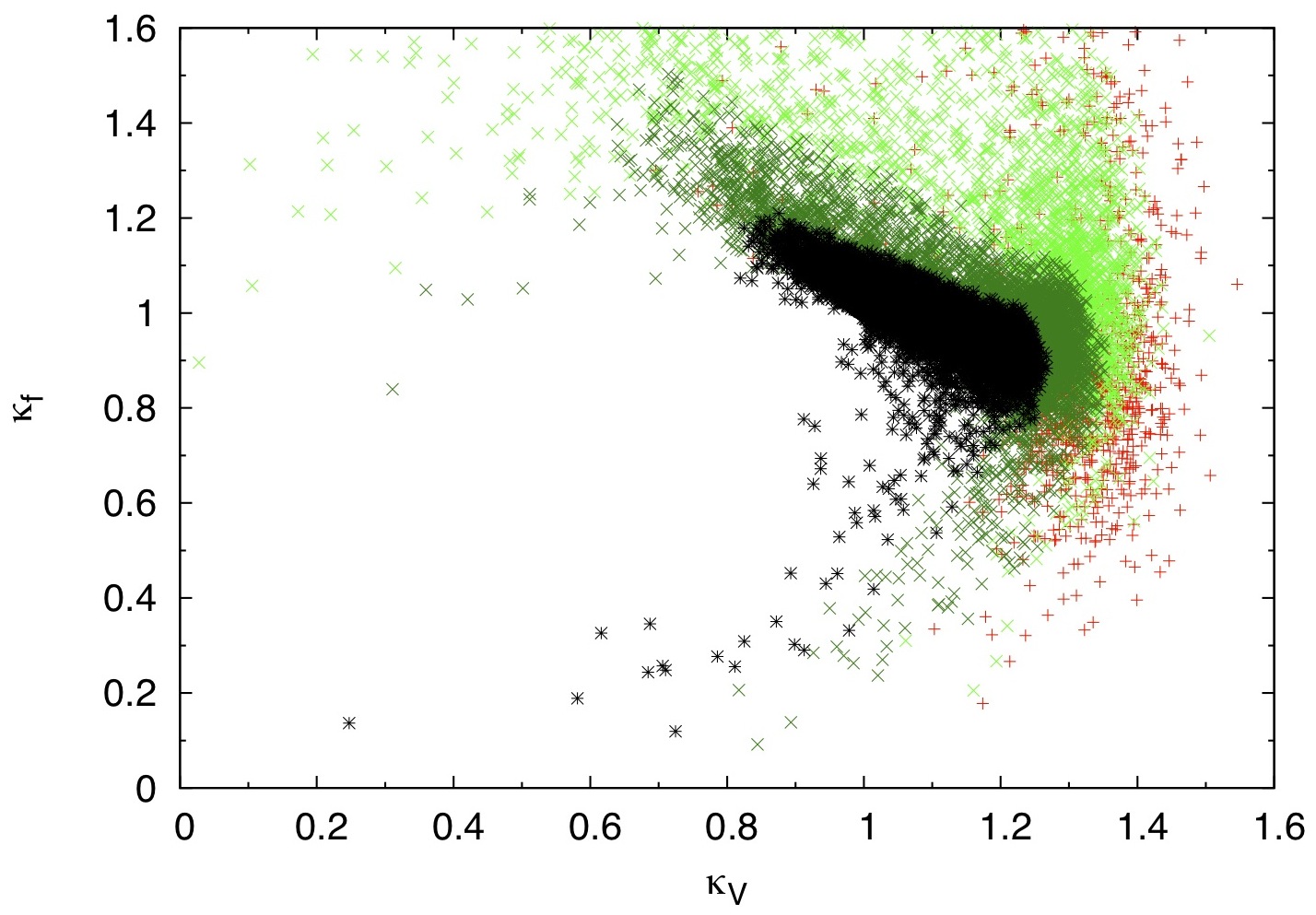}}
\caption{The allowed correlations between $\kappa_V$ and $\kappa_f$ after applying the constraints from ${\rm BR}(b \to s\gamma)$ and the $S$ parameter.  The color codes are the same as in Fig.~\ref{fig:mvchi}.}
\label{fig:kvkfcor}
\end{figure}

Crucially, however, the simultaneous enhancement of $\kappa_V$ and $\kappa_f$ occurs only when the new scalars are relatively light.  This is illustrated in Fig.~\ref{fig:kfkVmass}, where we plot $\kappa_V$ as a function of the mass of the lightest new scalar, for $\kappa_f$ within 5\% (red) or 10\% (blue) of $\kappa_V$.  The remaining points are shown in green.  Under the ``loose'' $b \to s \gamma$ constraint, for $\kappa_f$ within 5\% of $\kappa_V$, an 18\% enhancement of these couplings is possible only when at least one of the new scalars has mass below about 375~GeV.  This provides a complementary (albeit model dependent) way to constrain the flat direction by directly searching for the new scalars.  We leave a full consideration of the direct-search constraints on these additional scalars to future work.

\begin{figure}
\resizebox{0.5\textwidth}{!}{\includegraphics{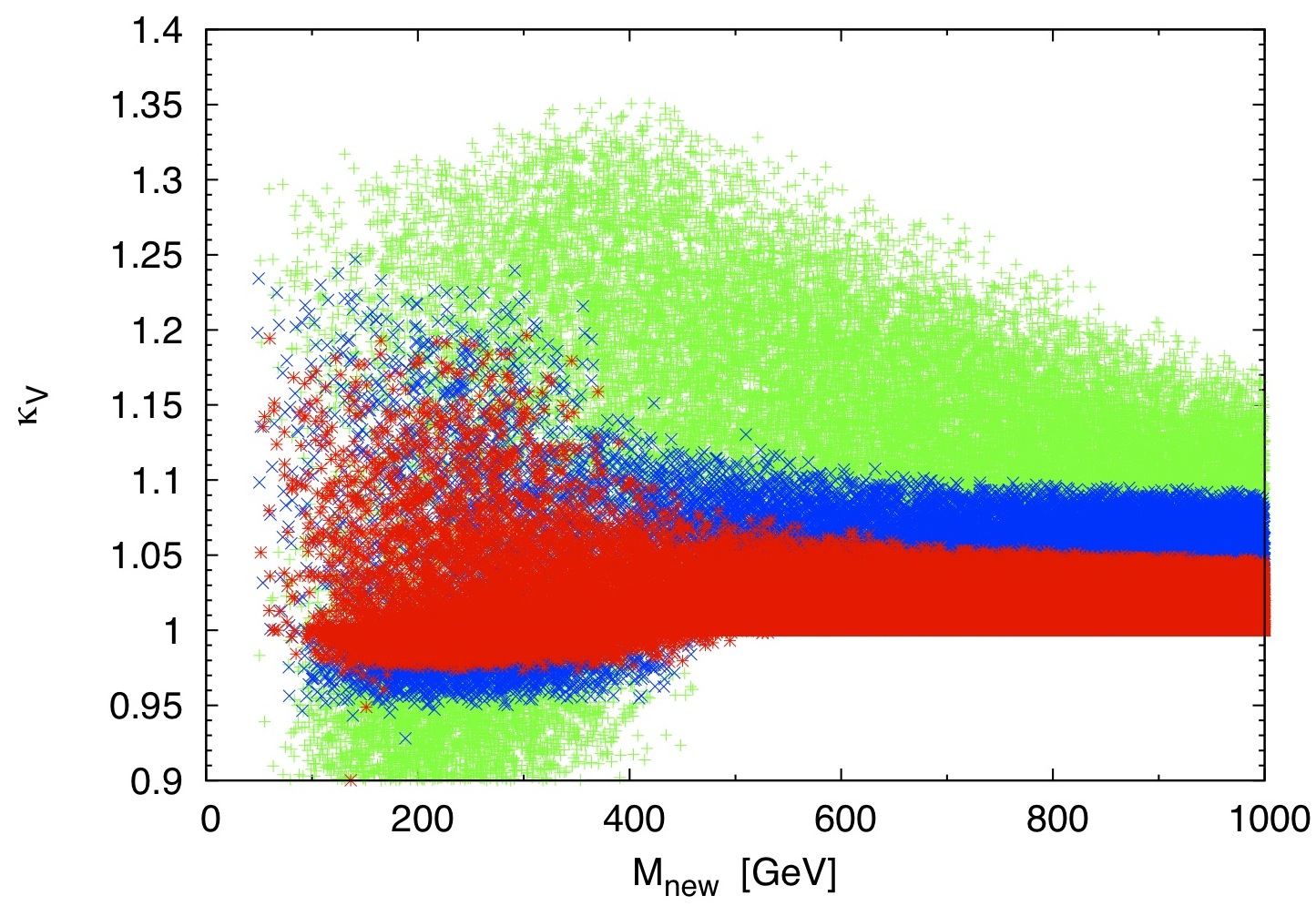}}%
\resizebox{0.5\textwidth}{!}{\includegraphics{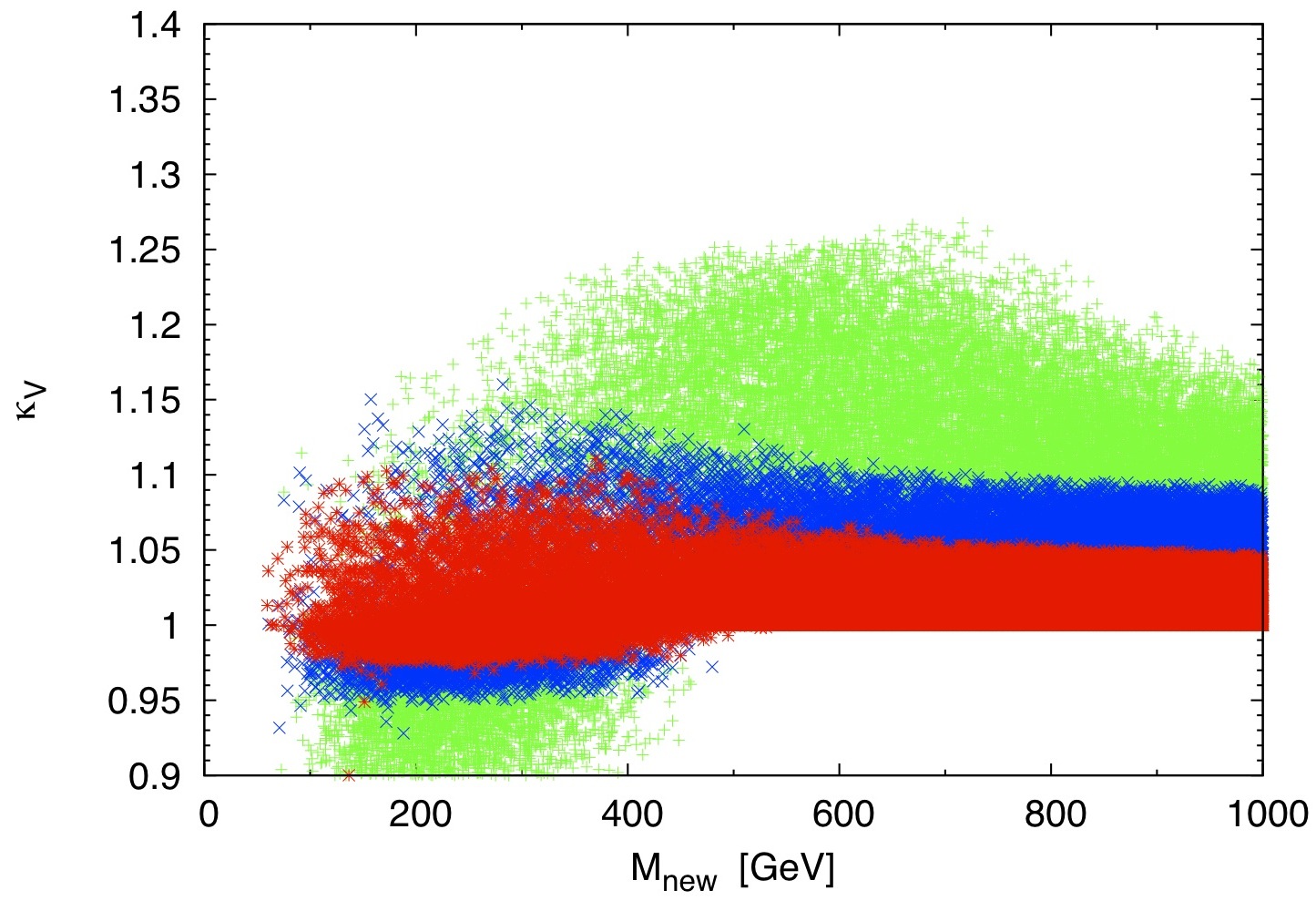}}
\caption{$\kappa_V$ as a function of the mass of the lightest new scalar, after imposing the constraint from the $S$ parameter and the ``loose'' (left) and ``tight'' (right) constraint from $b \to s \gamma$.  Points for which $|\kappa_f/\kappa_V - 1| < 5\%$ are shown in red (medium gray), points for which $|\kappa_f/\kappa_V - 1| < 10\%$ are shown in blue (dark gray), and the remaining points are shown in green (light gray).}
\label{fig:kfkVmass}
\end{figure}

\section{Conclusions}
\label{sec:conclusions}

In this paper we updated the indirect experimental constraints on the GM model coming from electroweak and $B$-physics observables---in particular the $S$ parameter, $R_b$, $b \to s \gamma$, $B^0_s$--$\bar B^0_s$ mixing, and $B_s^0 \to \mu^+\mu^-$.  Except for the $S$ parameter, all of these constrain only two of the GM model parameters: the isospin-triplet vev $v_{\chi}$ and the custodial-triplet mass $m_3$.  We gave the analytic expressions for the one-loop contributions from the additional Higgs bosons for the $S$ parameter, $R_b$, $B^0_s$--$\bar B^0_s$ mixing, and $B_s^0 \to \mu^+\mu^-$; in the case of the $S$ parameter and $R_b$ these are in the approximation that the new scalars are heavy compared to $M_Z$.  For $b \to s \gamma$ we adapted the 2HDM calculation implemented in the code SuperIso.  The constraints from $b \to s \gamma$, $B^0_s$--$\bar B^0_s$ mixing, and $B_s^0 \to \mu^+ \mu^-$ have not been studied in the GM model before.  

We found that $b \to s \gamma$ is currently the strongest of the $B$-physics constraints on the GM model.  However, this may be surpassed in the next few years by the constraint from $B_s^0 \to \mu^+\mu^-$, which will become more important as its statistical uncertainty is reduced with further LHC data-taking.  Combined with the theoretical requirements of vacuum stability and perturbativity, the $b\to s \gamma$ constraint puts a conservative upper bound of about 65~GeV on the isospin-triplet vev $v_{\chi}$, which leads to upper bounds on the $hWW$, $hZZ$, and $h f \bar f$ couplings.  In particular, a \emph{simultaneous} enhancement of the $hWW$, $hZZ$, and $h f \bar f$ couplings of up to 18\% compared to their SM values is still allowed by the indirect constraints, leading to a simultaneous enhancement of all the Higgs production cross sections by up to 39\%.  Such an enhancement could mask (and be masked by) the presence of undetected new decay modes of the SM-like Higgs boson at the LHC.

\begin{acknowledgments}
We thank F.~Mahmoudi for helpful discussions about SuperIso.  
H.E.L.\ also thanks the organizers of the Lisbon Workshop on Multi-Higgs Models 2014 for partial support and a stimulating environment while part of this work was performed.
This work was supported by the Natural Sciences and Engineering Research Council of Canada.  K.H.\ was also supported by the Government of Ontario through an Ontario Graduate Scholarship.
\end{acknowledgments}



\end{document}